# Proper identification of variable star V683 Cas

CORSATO SAMUELE[1], DI LAURO CHIARA[1], GASPERI GIULIA[1],
GRASSO SABRINA[1], INNOCENTE IRENE[1] AND POSSEMATO ROBERTA[1]
BENNA CARLO[2], GARDIOL DANIELE[2] AND PETTITI GIUSEPPE[2]
NESCI ROBERTO[3]

1) IIS Curie Vittorini, Corso Allamano 130, 10095, Grugliasco (TO), Italy, TOIS03400P@istruzione.it

2) INAF-Osservatorio Astrofisico di Torino, via Osservatorio 20, I-10025 Pino Torinese (TO), Italy, pettitg@alice.it

3) INAF-IAPS, via Fosso del Cavaliere 100, 00133 Roma (RM), Italy, Roberto.Nesci@iaps.inaf.it

**Abstract:** investigation on the object reported by SIMBAD database as V683 Cas shows that this variable star corresponds to the eclipsing binary system GSC 03699-00091 (also identified as NSVS 1850718). The results of the photometric observations of the stars nearby to the position indicated for V683 Cas are reported.

## 1 Introduction

This report describes our investigations and photometric observations performed on V683 Cas, that is part of a group of variable stars under investigation at the INAF-Astrophysical Observatory of Torino (OATo).

The object is reported as M303 and identified as an irregular variable star in the original study, the IBVS 3573 (Gasperoni 1991). The current identification of the variable star is defined in the IBVS 3840 (Kazarovets 1993).

This study revisits the original data of IBVS 3573, provides results of the photometric observations carried on at Loiano Astronomical Observatory and assesses data available from ASAS-SN database relevant to V683 Cas.

## 2 Original data

At the beginning of our investigations (October 2017), as shown in Figure 1, the position of V683 Cas shown by SIMBAD database did not match with any visible star.

Two different variable stars, separated by around 1', were indicated at the following coordinates:

- V683 Cas: R.A. = 02h 39m 03.0s, Dec. = +59° 43' 54'' (J2000).
- NSVS 1850718 alias GSC 03699-00091 or 2MASS J02390088+5942550: R.A. = 02h 39m 00.887s, Dec. = +59° 42' 55.00'' (J2000).

The coordinates of V683 Cas are those indicated in the original study (R.A. = 02h 35m 17s, Dec. = +59° 31') transformed to J2000 epoch, while the coordinates of 2MASS J02390088+5942550 are derived from NSVS database.

In order to get more more information for our analysis on V683 Cas, we analyzed the original photographic plates (Figure 2 and 3) and photometric data (Appendix 1) which were provided by courtesy of R. Nesci.

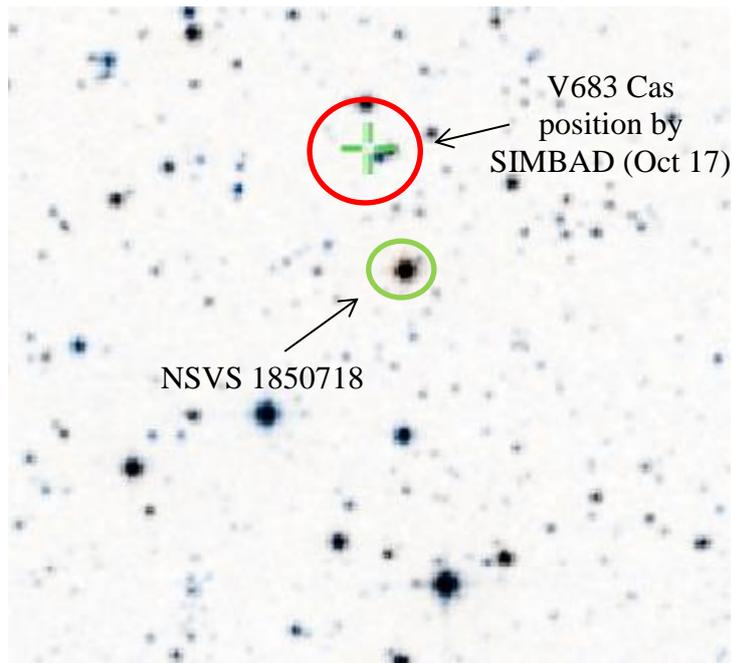

**Figure 1 – V683 Cas position and nearby variable star from SIMBAD (October 2017)**

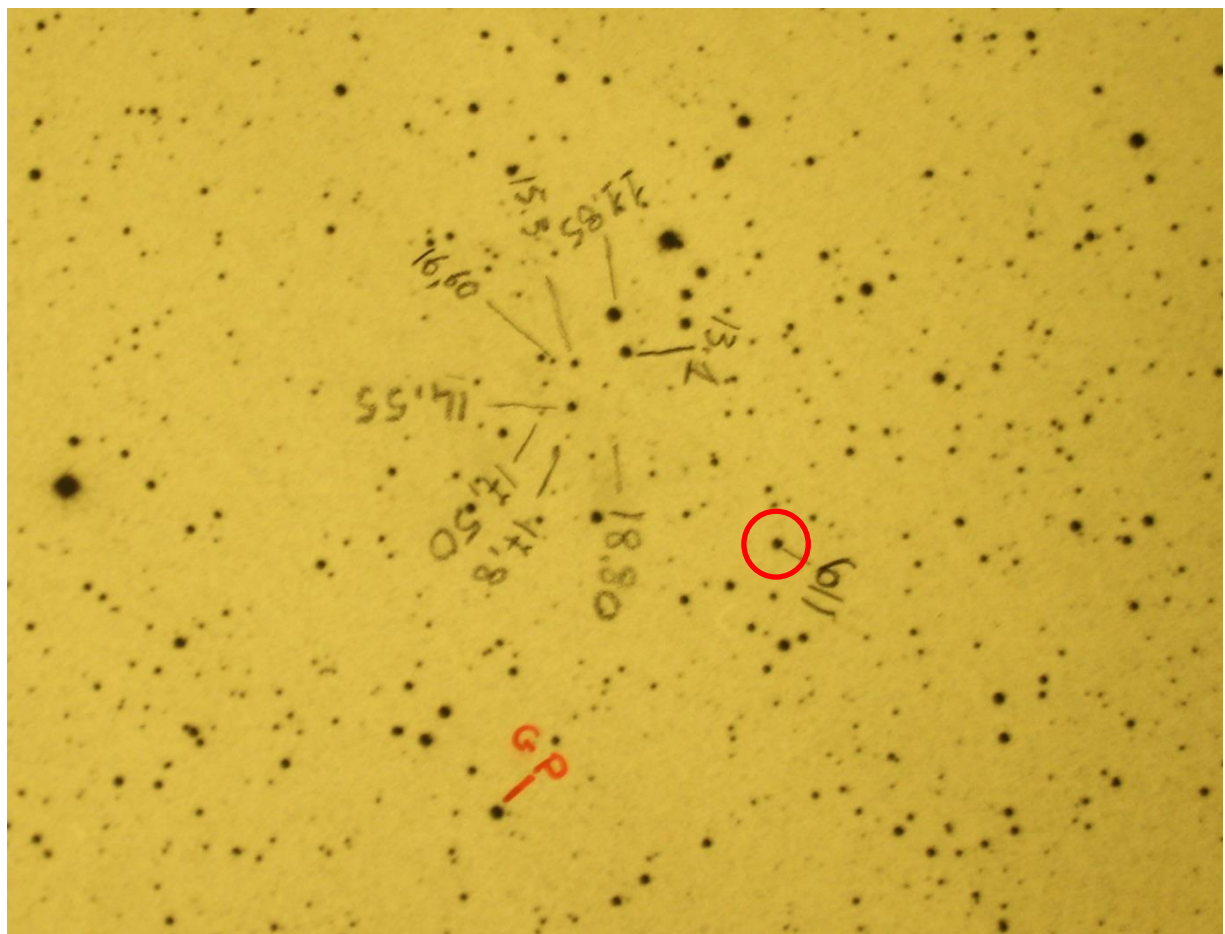

**Figure 2 - Photographic plate 103aO of M303 (V683 Cas)**

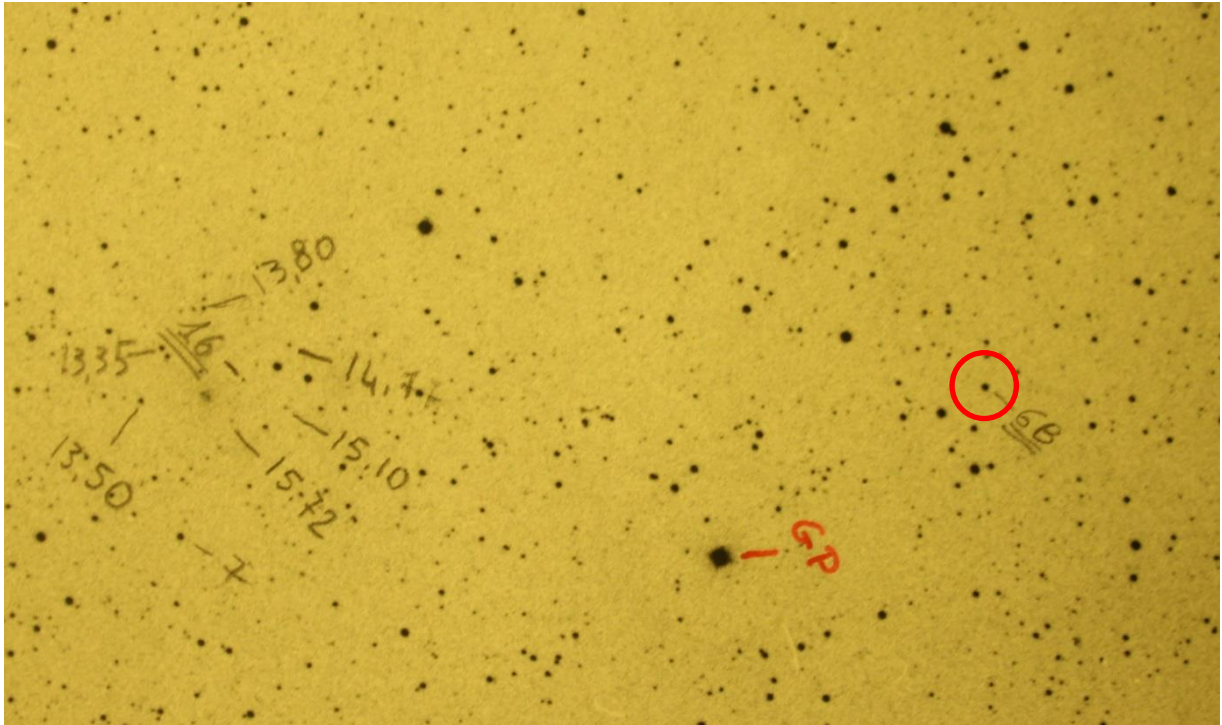

**Figure 3 - Photographic plate Kodak I-N plus RG5 filter of M303 (V683 Cas)**

Figures 2 and 3 show the variable star M303 (V683 Cas) labelled as ´6´ or ´6B´ respectively, together with the red variable star GP Cas and the comparison stars used for the photometry data reported in Appendix 1. The photographic plates in the blue band are 103aO, with or without GG13 filter, whilst the ones in the infrared are Kodak I-N plus RG5 filter. To a first approximation, the magnitudes, shown in Appendix 1, are comparable to the B and $I_c$ magnitudes of the Johnson-Cousins photometric standard system. The associated error is assumed to be $\pm 0.1^m$.

From the comparison of the original photographic plates with Figure 1 we found that the star identified with 6 or 6B corresponds to the star GSC 03699-00091 (also identified as NSVS 1850718 or 2MASS J02390088+5942550).

The mismatch in the coordinates is due to two different reasons:
1. The original M303 coordinates are provided with an error of 1'.
2. The SIMBAD database was referring to two different sources of data, when reporting the position of the two objects.

Previously to this report, the GSC 03699-00091 star was studied at OATo and shows a typical W UMa eclipsing binary light curve (Pettiti et al. in preparation); its light curve is shown in Figures 4 and 5, together with the M303 photometric data of IBVS 3573. No magnitude transformation was applied to the data of IBVS 3573. The magnitudes and the light curve trend of the IBVS 3573 data are comparable to the OATo ones.

The results of the photometric study on GSC 03699-00091 star, is being published (Pettiti et al. in preparation).

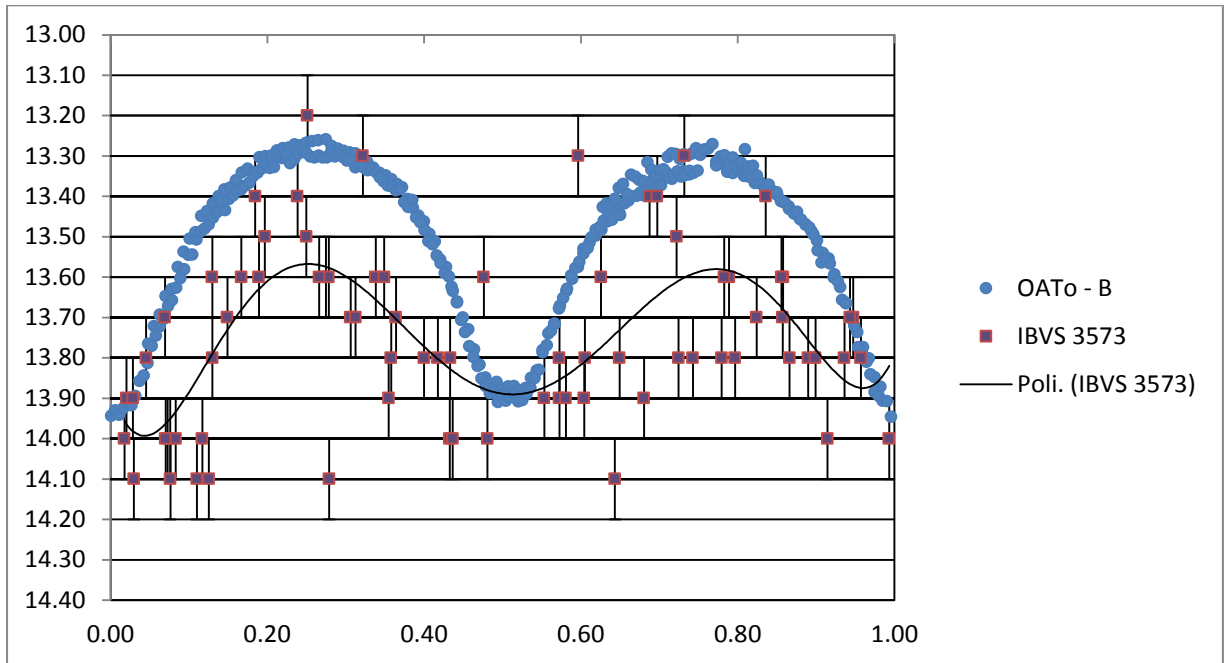

**Figure 4 – Light curve of GSC 03699-00091 in B band**

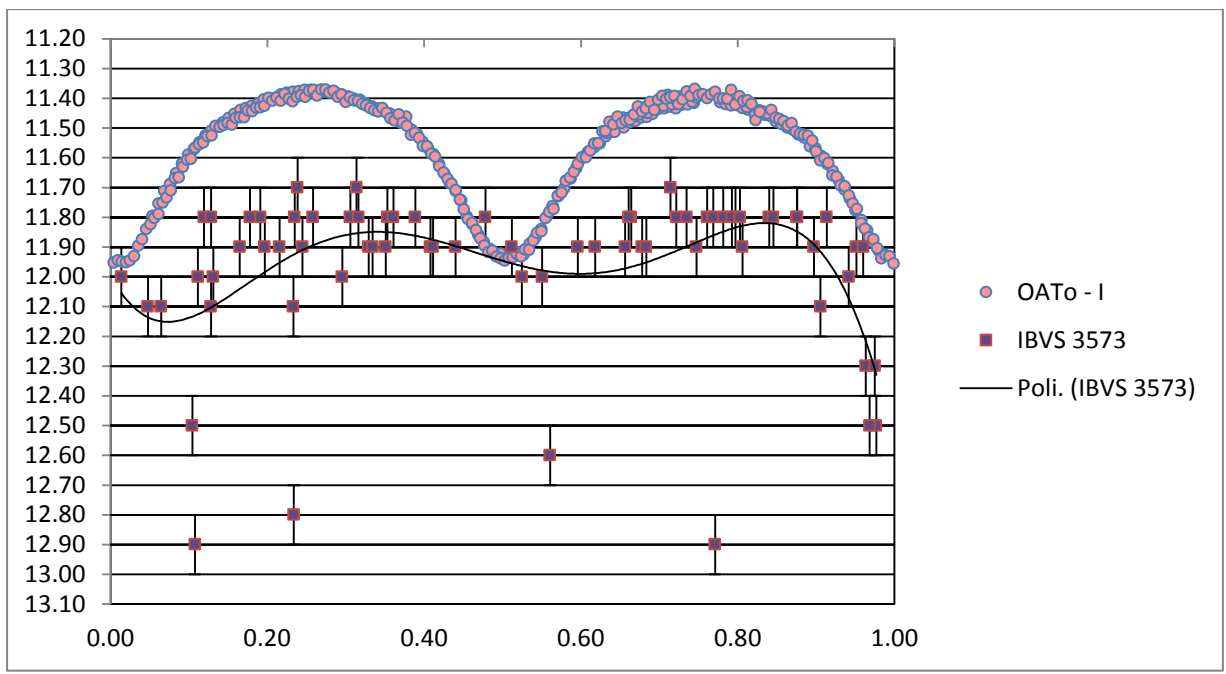

**Figure 5 - Light curve of GSC 03699-00091 in I band**

To check the presence of other possible variable stars nearby the position that SIMBAD was indicating for the object V683 Cas, we performed photometric observations of the following stars:

A. GSC 2.3 NAWK022826, Gaia DR1 464387841323087232 (R.A.=02h 39m 02.4639s, Dec. = +59° 43' 48.908"), J2000.
B. GSC 2.3 NAWK000139, Gaia DR1 464387841322326272 (R.A.=02h 39m 01.7460s, Dec. = +59° 43' 52.058"), J2000.

The stars A and B are the stars closest to the original coordinates indicated for V683 Cas and their position is shown in Figure 6.

The potential variability of stars 'A' and 'B' was also checked using data taken from the ASAS-SN database (Shappee et al. 2014, Kochanek et al. 2017).

The details of the photometric observations are discussed in section 3.

## 3  Observations

The photometric observations were carried out at Loiano site of the INAF-Bologna Astronomical Observatory (OABo).

The instrumentation characteristics are reported in Table 1.

| Loiano site of the Bologna Astronomical Observatory (OABo) | | | | | | |
|---|---|---|---|---|---|---|
| **Telescope (Cassini)** | | | **Detector (BFOSC)** | | | |
| **Useful Diameter** | **Focal Length** | **Optical conf.** | **Camera** | **Array (pixels)** | **Johnson-Kron-Cousins** | **FoV** |
| 150.0 cm | 1200 cm | Ritchey-Chrétien | EEV D129915 | 1300x1340 | B, V, R, I | 13'x12.6' |

**Table 1 - Loiano site instrumentation characteristics**

The number of observations and length of exposures in each filter is shown in Table 2.

| | | Number of observations - Length of exposure (s) | | | |
|---|---|---|---|---|---|
| **Date** | **Time span** | **B** | **V** | **R** | **I** |
| 7 Dec, 2016 | 7h 25min | 124-15s | 124-10s | 123-5s | 30-5s   93-4s |

**Table 2 - Observation log**

Data of the comparison star and check star are reported in Table 3. A finding chart with the identification of the stars is in Figure 6.

| **Star** | **ID** | *R.A. [h m s]*<br>*(J2000.0)* | *Dec. [° ´ ´´]*<br>*(J2000.0)* | *B* | *V* | *R* | *I* |
|---|---|---|---|---|---|---|---|
| Comparison<br>(C) | GSC 2.3<br>NAWK022754<br><br>Gaia DR1<br>464387772603607808 | 02 39 15.5936 | +59 43 44.319 | 17.61±0.03 | 15.90±0.04 | 15.15±0.03 | 14.38±0.04 |
| Check<br>(K) | GSC 2.3<br>NAWK022874<br><br>Gaia DR1<br>464387910042566144 | 02 38 59.2292 | +59 44 00.052 | 17.60±0.05 | 16.12±0.03 | 15.49±0.07 | 14.78±0.03 |

**Table 3 - Comparison and check star data**

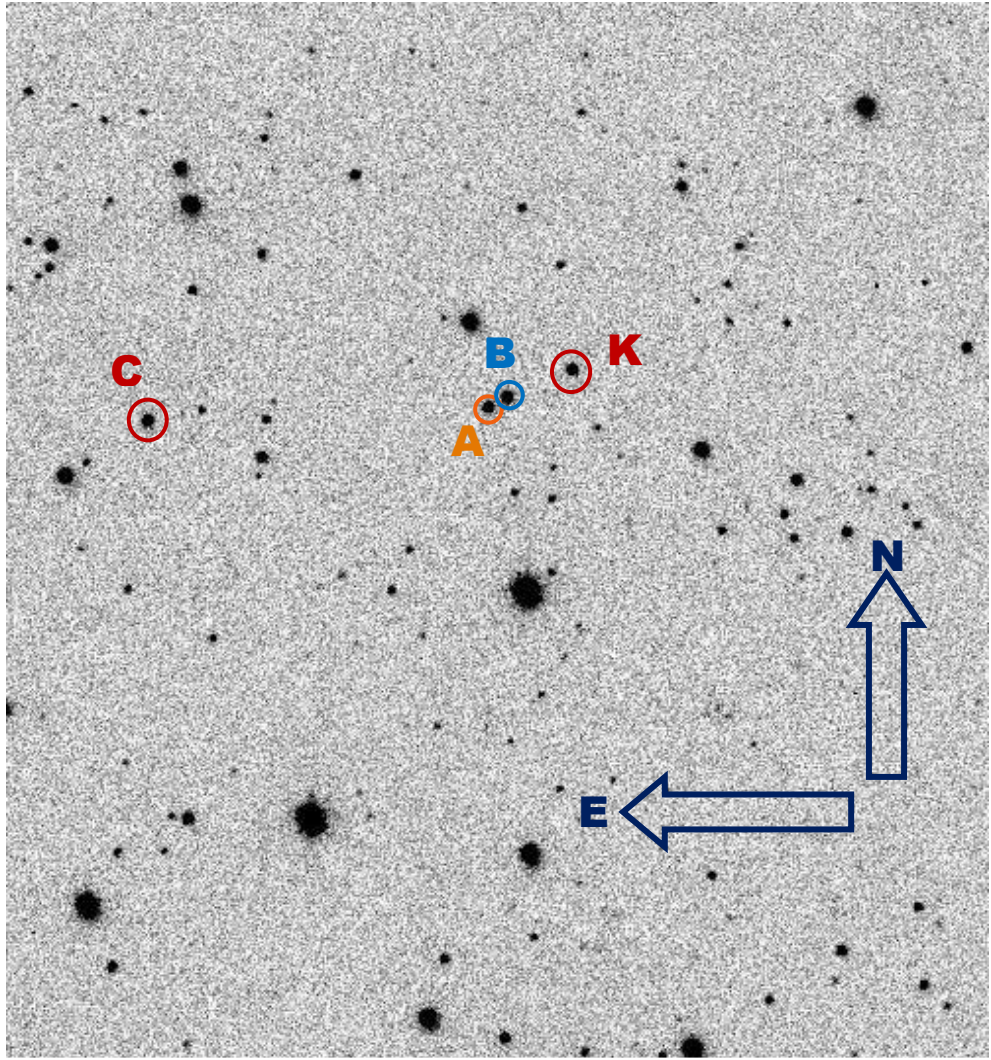

**Figure 6 – Finding chart for stars A and B**

## 4   Data analysis

For all images, Dark and Flat Field corrections were applied and aperture differential photometry was performed using the software AstroArt 5.0.

We used the basic equation to obtain standard magnitudes from instrumental magnitudes in the form (for filter V) shown in Eq. 1:

$$V_{var} = \Delta v + T_v * \Delta(B-V) + V_{comp} \qquad (1)$$

where:
- $\Delta v$ is the instrumental magnitude of the variable minus the instrumental magnitude of the comparison star;
- $V_{comp}$ is the V–magnitude of the comparison star defined in Table 3;
- $T_v$ is the transformation coefficient defined in Table 4;
- $\Delta(B-V)$ is the difference between the standard color of the variable and the standard color of the comparison star, computed using the formula:

$$\Delta(B-V) = T_{bv} * \Delta(b-v)$$

being $\Delta(b-v)$ the difference between the instrumental color of the variable and the instrumental color of the comparison star.

Similar equations were applied for filters B, $R_c$, $I_c$.

|  |  | **Loiano observations** |
|---|---|---|
| $T_b$ |  | 0.188 ± 0.013 |
| $T_v$ | for B, V | -0.034 ± 0.007 |
|  | for $R_c$ | -0.067 ± 0.015 |
|  | for $I_c$ | -0.034 ± 0.008 |
| $T_r$ |  | -0.060 ± 0.014 |
| $T_i$ |  | 0.033 ± 0.007 |
| $T_{bv}$ |  | 1.284 ± 0.008 |
| $T_{vr}$ |  | 1.000 ± 0.011 |
| $T_{vi}$ |  | 0.938 ± 0.008 |

**Table 4 - Transformation coefficients**

Observed data falling outside the range of ±3σ from the mean values have not been taken into account in the data analysis. The observed C-K magnitudes between the comparison and the check stars do not show significant trends or discontinuities and the error associated to the transformed standard magnitudes is dominated by the uncertainty of the comparison star.

## 5   Results

From our investigation on variable star M303 of IBVS 3573, we found that V683 Cas, labelled as a suspected irregular variable star, corresponds to eclipsing binary system GSC 03699-00091 (also identified as NSVS 1850718 or 2MASS J02390088+5942550).

To confirm this conclusion we analyzed photometric data of the stars Gaia DR1 464387841323087232 and Gaia DR1 464387841322326272, that are the stars closest to the original coordinates indicated for V683 Cas.

We did not find short or long term variability for these stars. For the short term analyses we used the observations performed to Loiano site, while for the long term analyses we utilized the data downloaded from ASAS-SN database. The search was carried out using the version 2.51 of the light curve and period analysis software PERANSO (Paunzen, E., Vanmunster, T. 2016).

The light curve of both stars in V filter is shown in Figures 7 and 8.

The mean values of the observed B, V, $R_c$, $I_c$ magnitudes and of the color indexes B-V, V-$R_c$, V-$I_c$ are summarized in Table 5. The infrared photometric characteristics of the two stars derived from the SIMBAD astronomical database (Cutri et al. 2003) are reported in Table 6. The color indexes V-J, V-H and V-K are calculated using the V mean values of Table 5.

All the measured BVRI photometric data, including the outliers, are reported in Appendix 2 and Appendix 3.

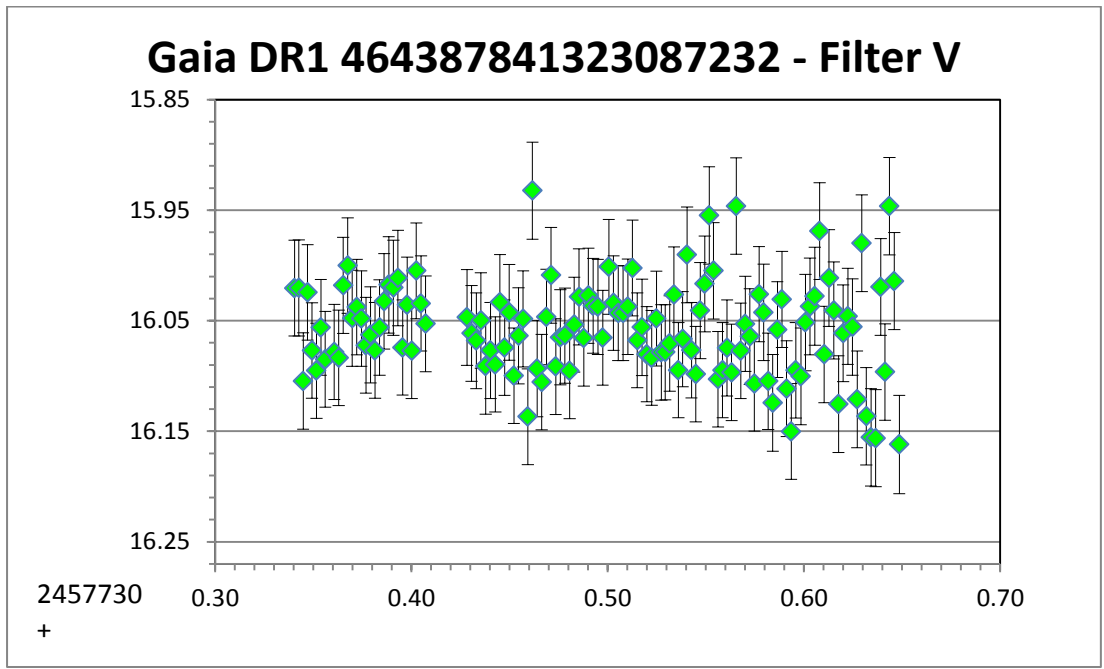

**Figure 7 – Light curve in V of Gaia DR1 464387841323087232**

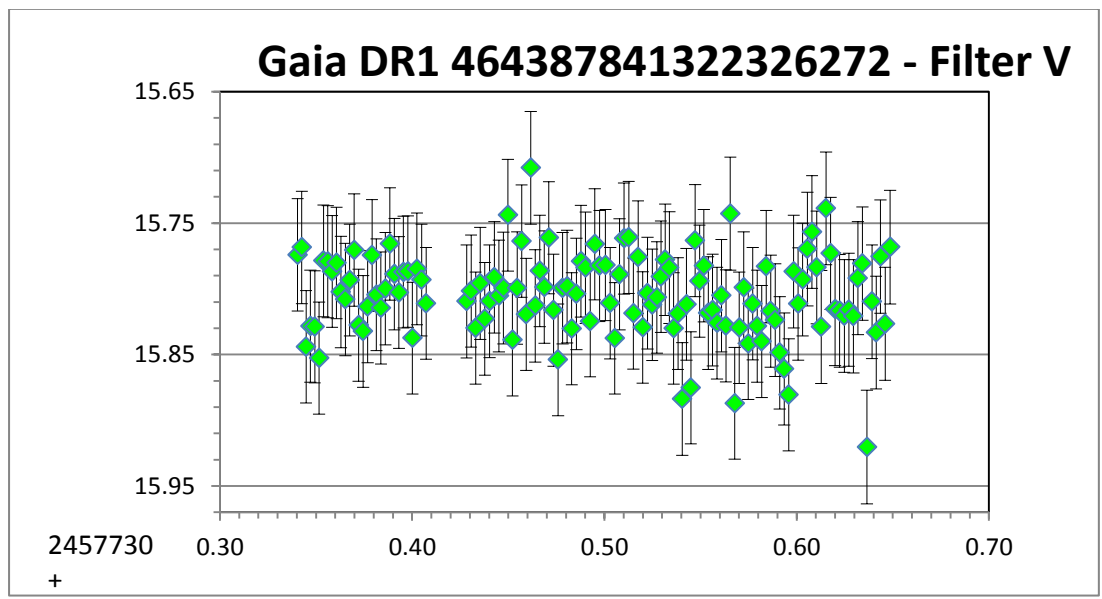

**Figure 8 – Light curve in V of Gaia DR1 464387841322326272**

|  | Gaia DR1 464387841323087232 | Gaia DR1 464387841322326272 |
|---|---|---|
| **B** | 18.39 ± 0.29 | 17.17 ± 0.22 |
| **V** | 16.06 ± 0.13 | 15.81 ± 0.10 |
| **R** | 15.01 ± 0.10 | 15.28 ± 0.10 |
| **I** | 13.95 ± 0.08 | 14.65 ± 0.08 |
| **B-V** | 2.33 ± 0.34 | 1.36 ± 0.26 |
| **V-R** | 1.05 ± 0.16 | 0.52 ± 0.14 |
| **V-I** | 2.11 ± 0.15 | 1.16 ± 0.10 |

**Table 5 – Mean values of BVRI photometric data (±3σ)**

|   | **Gaia DR1 464387841323087232** | **Gaia DR1 464387841322326272** |
|---|---|---|
| **J** | 12.449 ± 0.021 | 13.844 ± 0.038 |
| **H** | 11.714 ± 0.026 | 13.571 ± 0.043 |
| **K** | 11.492 ± 0.021 | 13.371 ± 0.035 |
| **J-H** | 0.735 ± 0.033 | 0.273 ± 0.057 |
| **H-K** | 0.222 ± 0.033 | 0.200 ± 0.055 |
| **V-J** | 3.61 ± 0.13 | 1.97 ± 0.11 |
| **V-H** | 4.35 ± 0.13 | 2.24 ± 0.11 |
| **V-K** | 4.57 ± 0.13 | 2.44 ± 0.11 |

**Table 6 – Infrared photometric characteristics**

## 6 Conclusions

We did not identify variable stars close to the original position indicated for V683 Cas. This report provides evidence that V683 Cas and the binary system GSC 03699-00091 (also identified as NSVS 1850718 or 2MASS J02390088+5942550), registered as different sources with respect to position and variability type, are in fact the same source. The most precise position for this object is R.A. 02h 39m 00.987s, DEC. +59° 42' 55.30" (ICRS J2000), as available from Gaia DR1, and the confirmed variability type is W UMa.

## Acknowledgements

- We acknowledge the use of the 1.52m Cassini Telescope run by INAF-Osservatorio Astronomico di Bologna at Loiano site.
- This activity has made use of the SIMBAD database, operated at CDS, Strasbourg, France.
- This work has made use of data from the European Space Agency (ESA) mission Gaia (https://www.cosmos.esa.int/gaia), processed by the Gaia Data Processing and Analysis Consortium (DPAC, https://www.cosmos.esa.int/web/gaia/dpac/consortium). Funding for the DPAC has been provided by national institutions, in particular the institutions participating in the Gaia Multilateral Agreement.
- This work was carried out in the context of educational and training activities provided by Italian law 'Alternanza Scuola Lavoro', July 13th, 2015 n.107, Art.1, paragraphs 33-43.

# Appendix 1
# M303 photometric data

| M303 – Magnitude B | | | | |
|---|---|---|---|---|
| **Date** | **JD** | **HJD** | **Magnitude B** | **Mag. B Error (±)** |
| 4-nov-62 | 2437973.51299 | 2437973.51709 | 13.4 | 0.1 |
| 4-nov-62 | 2437973.51299 | 2437973.51709 | 13.4 | 0.1 |
| 6-nov-62 | 2437975.49983 | 2437975.50397 | 14.0 | 0.1 |
| 22-mar-63 | 2438111.33291 | 2438111.33058 | 13.3 | 0.1 |
| 12-oct-63 | 2438315.5 | 2438315.5 | 13.6 | 0.1 |
| 29-sep-67 | 2439763.51651 | 2439763.51914 | 13.8 | 0.1 |
| 1-oct-67 | 2439765.49996 | 2439765.50270 | 13.6 | 0.1 |
| 6-oct-67 | 2439770.48082 | 2439770.48383 | 13.8 | 0.1 |
| 8-oct-67 | 2439772.44411 | 2439772.44722 | 13.9 | 0.1 |
| 11-oct-67 | 2439775.57441 | 2439775.57768 | 13.7 | 0.1 |
| 2-dec-67 | 2439827.69779 | 2439827.70196 | 13.5 | 0.1 |
| 2-jan-68 | 2439858.55724 | 2439858.56035 | 14.0 | 0.1 |
| 4-jan-68 | 2439860.39265 | 2439860.39566 | 13.4 | 0.1 |
| 20-jan-68 | 2439876.36478 | 2439876.36683 | 13.4 | 0.1 |
| 23-jan-68 | 2439879.49418 | 2439879.49602 | 13.8 | 0.1 |
| 21-jul-68 | 2440059.53254 | 2440059.53038 | 13.6 | 0.1 |
| 22-jul-68 | 2440060.54128 | 2440060.53918 | 14.0 | 0.1 |
| 19-sep-68 | 2440119.62077 | 2440119.62284 | 13.8 | 0.1 |
| 27-sep-68 | 2440127.61136 | 2440127.61392 | 13.8 | 0.1 |
| 15-oct-68 | 2440145.31314 | 2440145.31661 | 13.6 | 0.1 |
| 19-oct-68 | 2440149.32300 | 2440149.32664 | 13.4 | 0.1 |
| 24-oct-68 | 2440154.28580 | 2440154.28961 | 13.2 | 0.1 |
| 21-dec-68 | 2440212.28786 | 2440212.29149 | 14.1 | 0.1 |
| 17-jan-69 | 2440239.31607 | 2440239.31827 | 13.3 | 0.1 |
| 21-jan-69 | 2440243.29361 | 2440243.29555 | 13.5 | 0.1 |
| 13-feb-69 | 2440266.25379 | 2440266.25408 | 13.7 | 0.1 |
| 21-feb-69 | 2440274.27266 | 2440274.27235 | 13.6 | 0.1 |
| 12-jul-69 | 2440415.56086 | 2440415.55814 | 13.7 | 0.1 |
| 13-sep-69 | 2440478.39877 | 2440478.40041 | 14.0 | 0.1 |
| 20-sep-69 | 2440485.66111 | 2440485.66323 | 13.8 | 0.1 |
| 22-sep-69 | 2440487.65228 | 2440487.65452 | 13.9 | 0.1 |
| 11-oct-69 | 2440506.41433 | 2440506.41761 | 13.7 | 0.1 |
| 15-oct-69 | 2440510.35967 | 2440510.36313 | 13.6 | 0.1 |
| 7-dec-69 | 2440563.34565 | 2440563.34972 | 13.5 | 0.1 |
| 2-jan-70 | 2440589.22661 | 2440589.22971 | 13.8 | 0.1 |
| 7-feb-70 | 2440625.43747 | 2440625.43821 | 13.4 | 0.1 |
| 1-mar-70 | 2440647.38861 | 2440647.38771 | 13.6 | 0.1 |
| 4-aug-70 | 2440803.57997 | 2440803.57872 | 13.8 | 0.1 |
| 6-aug-70 | 2440805.53775 | 2440805.53664 | 13.6 | 0.1 |
| 24-aug-70 | 2440823.42796 | 2440823.42815 | 13.6 | 0.1 |
| 27-oct-70 | 2440887.36352 | 2440887.36742 | 13.8 | 0.1 |
| 28-jan-71 | 2440980.27096 | 2440980.27246 | 13.8 | 0.1 |
| 21-feb-71 | 2441004.29210 | 2441004.29182 | 13.6 | 0.1 |
| 28-feb-71 | 2441011.35682 | 2441011.35602 | 13.7 | 0.1 |
| 29-mar-71 | 2441040.32056 | 2441040.31781 | 14.0 | 0.1 |
| 15-apr-71 | 2441057.32083 | 2441057.31722 | 13.6 | 0.1 |
| 30-jul-71 | 2441163.42893 | 2441163.42731 | 13.8 | 0.1 |
| 16-aug-71 | 2441180.52456 | 2441180.52416 | 14.0 | 0.1 |
| 19-aug-71 | 2441183.47964 | 2441183.47945 | 13.8 | 0.1 |
| 21-sep-71 | 2441216.61742 | 2441216.61957 | 13.9 | 0.1 |
| 24-oct-71 | 2441249.27454 | 2441249.27833 | 13.8 | 0.1 |

| M303 – Magnitude B ||||| 
|---|---|---|---|---|
| **Date** | **JD** | **HJD** | **Magnitude B** | **Mag. B Error (±)** |
| 17-nov-71 | 2441273.47294 | 2441273.47720 | 13.8 | 0.1 |
| 15-jan-72 | 2441332.23008 | 2441332.23246 | 13.8 | 0.1 |
| 16-aug-72 | 2441546.53854 | 2441546.53819 | 13.9 | 0.1 |
| 25-feb-73 | 2441739.42373 | 2441739.42310 | 14.0 | 0.1 |
| 31-jul-73 | 2441895.56372 | 2441895.56221 | 13.8 | 0.1 |
| 4-sep-73 | 2441930.51002 | 2441930.51104 | 13.7 | 0.1 |
| 5-sep-73 | 2441931.58370 | 2441931.58480 | 13.7 | 0.1 |
| 8-sep-73 | 2441934.63183 | 2441934.63314 | 13.8 | 0.1 |
| 27-sep-73 | 2441953.48097 | 2441953.48351 | 13.7 | 0.1 |
| 28-sep-73 | 2441954.41089 | 2441954.41348 | 13.6 | 0.1 |
| 4-oct-73 | 2441960.55570 | 2441960.55864 | 14.0 | 0.1 |
| 24-oct-73 | 2441980.51760 | 2441980.52142 | 13.9 | 0.1 |
| 21-jan-74 | 2442069.32347 | 2442069.32542 | 14.0 | 0.1 |
| 14-mar-74 | 2442121.29395 | 2442121.29213 | 13.7 | 0.1 |
| 11-sep-74 | 2442302.59794 | 2442302.59944 | 14.1 | 0.1 |
| 13-sep-74 | 2442304.52794 | 2442304.52957 | 14.1 | 0.1 |
| 15-sep-74 | 2442306.28850 | 2442306.29025 | 14.1 | 0.1 |
| 26-sep-74 | 2442317.60452 | 2442317.60699 | 14.1 | 0.1 |
| 22-oct-74 | 2442343.55183 | 2442343.55557 | 13.8 | 0.1 |
| 7-dec-74 | 2442389.38731 | 2442389.39139 | 14.0 | 0.1 |
| 19-dec-74 | 2442401.41687 | 2442401.42059 | 13.6 | 0.1 |
| 7-jan-75 | 2442420.41545 | 2442420.41828 | 13.3 | 0.1 |
| 6-feb-75 | 2442450.33070 | 2442450.33154 | 14.0 | 0.1 |
| 3-sep-75 | 2442659.57870 | 2442659.57962 | 13.6 | 0.1 |
| 7-sep-75 | 2442663.46252 | 2442663.46371 | 13.9 | 0.1 |
| 2-oct-75 | 2442688.52532 | 2442688.52812 | 13.9 | 0.1 |
| 26-dec-84 | 2446061.44731 | 2446061.45072 | 13.9 | 0.1 |
| 25-jan-85 | 2446091.41837 | 2446091.42002 | 14.1 | 0.1 |

| M303 - Magnitude I ||||| 
|---|---|---|---|---|
| **Date** | **JD** | **HJD** | **Magnitude I** | **Mag. I Error (±)** |
| 29-sep-67 | 2439763.54568 | 2439763.54831 | 11.8 | 0.1 |
| 1-oct-67 | 2439765.55171 | 2439765.55445 | 11.9 | 0.1 |
| 2-jan-68 | 2439858.63008 | 2439858.63318 | 11.8 | 0.1 |
| 4-jan-68 | 2439860.30359 | 2439860.30661 | 12.1 | 0.1 |
| 20-jan-68 | 2439876.36478 | 2439876.36683 | 11.7 | 0.1 |
| 23-jan-68 | 2439879.51780 | 2439879.51964 | 11.8 | 0.1 |
| 5-jul-68 | 2440043.55454 | 2440043.55145 | 11.9 | 0.1 |
| 21-jul-68 | 2440059.55754 | 2440059.55538 | 11.8 | 0.1 |
| 22-jul-68 | 2440060.59219 | 2440060.59010 | 11.9 | 0.1 |
| 19-sep-68 | 2440119.59508 | 2440119.59714 | 12.0 | 0.1 |
| 27-sep-68 | 2440127.63444 | 2440127.63700 | 11.8 | 0.1 |
| 19-oct-68 | 2440149.33898 | 2440149.34262 | 11.9 | 0.1 |
| 24-oct-68 | 2440154.30578 | 2440154.30960 | 11.7 | 0.1 |
| 24-nov-68 | 2440185.41315 | 2440185.41740 | 11.9 | 0.1 |
| 21-dec-68 | 2440212.27096 | 2440212.27459 | 12.5 | 0.1 |
| 17-jan-69 | 2440239.32857 | 2440239.33077 | 12.9 | 0.1 |
| 21-jan-69 | 2440243.30541 | 2440243.30735 | 12.8 | 0.1 |
| 13-feb-69 | 2440266.24060 | 2440266.24089 | 12.9 | 0.1 |
| 21-feb-69 | 2440274.25877 | 2440274.25846 | 11.8 | 0.1 |
| 12-jul-69 | 2440415.54311 | 2440415.54039 | 12.0 | 0.1 |
| 13-sep-69 | 2440478.37447 | 2440478.37611 | 12.0 | 0.1 |
| 20-sep-69 | 2440485.64027 | 2440485.64239 | 11.7 | 0.1 |
| 22-sep-69 | 2440487.63414 | 2440487.63638 | 12.3 | 0.1 |

| M303 - Magnitude I ||||| 
|---|---|---|---|---|
| **Date** | **JD** | **HJD** | **Magnitude I** | **Mag. I Error (±)** |
| 11-oct-69 | 2440506.39423 | 2440506.39751 | 11.8 | 0.1 |
| 15-oct-69 | 2440510.34023 | 2440510.34369 | 11.8 | 0.1 |
| 16-nov-69 | 2440542.41596 | 2440542.42022 | 11.9 | 0.1 |
| 7-dec-69 | 2440563.32690 | 2440563.33097 | 11.8 | 0.1 |
| 2-jan-70 | 2440589.20994 | 2440589.21304 | 11.8 | 0.1 |
| 7-feb-70 | 2440625.42020 | 2440625.42094 | 11.8 | 0.1 |
| 1-mar-70 | 2440647.40529 | 2440647.40439 | 11.9 | 0.1 |
| 4-aug-70 | 2440803.59525 | 2440803.59401 | 11.8 | 0.1 |
| 6-aug-70 | 2440805.55789 | 2440805.55678 | 11.9 | 0.1 |
| 24-aug-70 | 2440823.45018 | 2440823.45038 | 11.9 | 0.1 |
| 27-oct-70 | 2440887.35033 | 2440887.35423 | 11.9 | 0.1 |
| 28-jan-71 | 2440980.29596 | 2440980.29745 | 12.5 | 0.1 |
| 21-feb-71 | 2441004.30758 | 2441004.30730 | 11.9 | 0.1 |
| 28-feb-71 | 2441011.34154 | 2441011.34074 | 11.8 | 0.1 |
| 28-mar-71 | 2441039.33207 | 2441039.32937 | 12.3 | 0.1 |
| 29-mar-71 | 2441040.33063 | 2441040.32787 | 12.5 | 0.1 |
| 15-apr-71 | 2441057.31611 | 2441057.31250 | 11.9 | 0.1 |
| 30-jul-71 | 2441163.44837 | 2441163.44675 | 11.9 | 0.1 |
| 13-aug-71 | 2441177.52018 | 2441177.51956 | 12.1 | 0.1 |
| 16-aug-71 | 2441180.54678 | 2441180.54638 | 12.0 | 0.1 |
| 19-aug-71 | 2441183.49700 | 2441183.49681 | 11.8 | 0.1 |
| 21-sep-71 | 2441216.60353 | 2441216.60568 | 12.6 | 0.1 |
| 24-oct-71 | 2441249.26065 | 2441249.26444 | 11.8 | 0.1 |
| 17-nov-71 | 2441273.48683 | 2441273.49109 | 11.8 | 0.1 |
| 15-jan-72 | 2441332.21758 | 2441332.21996 | 11.9 | 0.1 |
| 16-aug-72 | 2441546.55303 | 2441546.55268 | 11.9 | 0.1 |
| 6-jan-73 | 2441689.40141 | 2441689.40427 | 11.8 | 0.1 |
| 25-feb-73 | 2441739.44039 | 2441739.43976 | 12.1 | 0.1 |
| 31-jul-73 | 2441895.54566 | 2441895.54415 | 11.8 | 0.1 |
| 4-sep-73 | 2441930.48703 | 2441930.48805 | 11.8 | 0.1 |
| 5-sep-73 | 2441931.56210 | 2441931.56319 | 11.9 | 0.1 |
| 8-sep-73 | 2441934.48669 | 2441934.48799 | 11.9 | 0.1 |
| 27-sep-73 | 2441953.45944 | 2441953.46198 | 11.8 | 0.1 |
| 28-sep-73 | 2441954.39006 | 2441954.39265 | 11.8 | 0.1 |
| 4-oct-73 | 2441960.57098 | 2441960.57392 | 11.9 | 0.1 |
| 24-oct-73 | 2441980.53496 | 2441980.53878 | 11.8 | 0.1 |
| 1-dec-73 | 2442018.41099 | 2442018.41517 | 11.9 | 0.1 |
| 21-jan-74 | 2442069.34083 | 2442069.34278 | 12.1 | 0.1 |
| 14-mar-74 | 2442121.27659 | 2442121.27477 | 11.8 | 0.1 |
| 15-sep-74 | 2442306.56558 | 2442306.56735 | 11.9 | 0.1 |
| 22-oct-74 | 2442343.56615 | 2442343.56989 | 11.8 | 0.1 |
| 7-dec-74 | 2442389.40259 | 2442389.40667 | 12.0 | 0.1 |
| 19-dec-74 | 2442401.59812 | 2442401.60183 | 11.9 | 0.1 |
| 7-jan-75 | 2442420.43678 | 2442420.43961 | 11.8 | 0.1 |
| 3-sep-75 | 2442659.55092 | 2442659.55184 | 11.8 | 0.1 |
| 7-sep-75 | 2442663.48891 | 2442663.49011 | 12.0 | 0.1 |
| 2-oct-75 | 2442688.60974 | 2442688.61255 | 11.8 | 0.1 |
| 26-dec-84 | 2446061.42840 | 2446061.43181 | 12.0 | 0.1 |
| 25-jan-85 | 2446091.40378 | 2446091.40543 | 12.1 | 0.1 |

# Appendix 2
# BVR$_c$I$_c$ photometric data of star Gaia DR1 464387841323087232

| Gaia DR1 464387841323087232 ||||||||||||
|---|---|---|---|---|---|---|---|---|---|---|---|
| HJD (2400000 +) | B | Err. (±) | HJD (2400000 +) | V | Err. (±) | HJD (2400000 +) | R$_c$ | Err. (±) | HJD (2400000 +) | I$_c$ | Err. (±) |
| 57730.33978 | 18.23 | 0.05 | 57730.34039 | 16.02 | 0.04 | 57730.34088 | 15.00 | 0.03 | 57730.34139 | 13.97 | 0.04 |
| 57730.34199 | 18.47 | 0.05 | 57730.34260 | 16.02 | 0.04 | 57730.34309 | 15.00 | 0.03 | 57730.34361 | 13.93 | 0.04 |
| 57730.34421 | 18.46 | 0.05 | 57730.34482 | 16.10 | 0.04 | 57730.34532 | 15.05 | 0.03 | 57730.34585 | 13.98 | 0.04 |
| 57730.34646 | 18.34 | 0.05 | 57730.34707 | 16.02 | 0.04 | 57730.34757 | 15.02 | 0.03 | 57730.34809 | 14.00 | 0.04 |
| 57730.34870 | 18.22 | 0.05 | 57730.34931 | 16.08 | 0.04 | 57730.34982 | 14.98 | 0.03 | 57730.35035 | 13.94 | 0.04 |
| 57730.35095 | 18.33 | 0.05 | 57730.35157 | 16.10 | 0.04 | 57730.35207 | 14.98 | 0.03 | 57730.35260 | 13.94 | 0.04 |
| 57730.35322 | 18.34 | 0.05 | 57730.35383 | 16.06 | 0.04 | 57730.35434 | 14.99 | 0.03 | 57730.35486 | 13.92 | 0.04 |
| 57730.35547 | 18.30 | 0.05 | 57730.35609 | 16.08 | 0.04 | 57730.35660 | 15.02 | 0.03 | 57730.35713 | 13.97 | 0.04 |
| 57730.35776 | 18.55 | 0.05 | 57730.35837 | 15.69 | 0.04 | 57730.35888 | 15.05 | 0.03 | 57730.35941 | 13.94 | 0.04 |
| 57730.36002 | 18.28 | 0.05 | 57730.36065 | 16.08 | 0.04 | 57730.36115 | 15.03 | 0.03 | 57730.36169 | 13.96 | 0.04 |
| 57730.36232 | 18.37 | 0.05 | 57730.36293 | 16.08 | 0.04 | 57730.36239 | 15.02 | 0.03 | 57730.36398 | 13.96 | 0.04 |
| 57730.36461 | 18.45 | 0.05 | 57730.36524 | 16.02 | 0.04 | 57730.36574 | 15.03 | 0.03 | 57730.36627 | 13.95 | 0.04 |
| 57730.36690 | 18.46 | 0.05 | 57730.36752 | 16.00 | 0.04 | 57730.36804 | 15.01 | 0.03 | 57730.36858 | 13.92 | 0.04 |
| 57730.36920 | 18.34 | 0.05 | 57730.36982 | 16.05 | 0.04 | 57730.37034 | 14.98 | 0.03 | 57730.37088 | 13.96 | 0.04 |
| 57730.37152 | 18.39 | 0.05 | 57730.37215 | 16.04 | 0.04 | 57730.37265 | 15.01 | 0.03 | 57730.37321 | 13.91 | 0.04 |
| 57730.37383 | 18.45 | 0.05 | 57730.37446 | 16.05 | 0.04 | 57730.37498 | 15.01 | 0.03 | 57730.37552 | 13.92 | 0.04 |
| 57730.37615 | 18.45 | 0.05 | 57730.37679 | 16.07 | 0.04 | 57730.37730 | 15.02 | 0.03 | 57730.37784 | 13.92 | 0.04 |
| 57730.37847 | 18.53 | 0.05 | 57730.37910 | 16.06 | 0.04 | 57730.37962 | 15.03 | 0.03 | 57730.38015 | 13.93 | 0.04 |
| 57730.38079 | 18.34 | 0.05 | 57730.38141 | 16.08 | 0.04 | 57730.38193 | 14.99 | 0.03 | 57730.38248 | 13.98 | 0.04 |
| 57730.38309 | 18.35 | 0.05 | 57730.38373 | 16.06 | 0.04 | 57730.38426 | 14.99 | 0.03 | 57730.38481 | 13.96 | 0.04 |
| 57730.38544 | 18.23 | 0.05 | 57730.38608 | 16.03 | 0.04 | 57730.38661 | 14.98 | 0.03 | 57730.38716 | 13.95 | 0.04 |
| 57730.38780 | 18.39 | 0.05 | 57730.38843 | 16.02 | 0.04 | 57730.38896 | 15.04 | 0.03 | 57730.38951 | 13.97 | 0.04 |
| 57730.39015 | 18.36 | 0.05 | 57730.39079 | 16.02 | 0.04 | 57730.39131 | 15.05 | 0.03 | 57730.39185 | 13.93 | 0.04 |
| 57730.39249 | 18.45 | 0.05 | 57730.39313 | 16.01 | 0.04 | 57730.39366 | 15.00 | 0.03 | 57730.39420 | 13.96 | 0.04 |
| 57730.39483 | 18.21 | 0.05 | 57730.39546 | 16.07 | 0.04 | 57730.39598 | 15.04 | 0.03 | 57730.39653 | 13.94 | 0.04 |
| 57730.39715 | 18.40 | 0.05 | 57730.39778 | 16.04 | 0.04 | 57730.39831 | 14.97 | 0.03 | 57730.39885 | 13.96 | 0.04 |
| 57730.39949 | 18.48 | 0.05 | 57730.40013 | 16.08 | 0.04 | 57730.40066 | 14.98 | 0.03 | 57730.40120 | 13.98 | 0.04 |
| 57730.40184 | 18.55 | 0.05 | 57730.40248 | 16.00 | 0.04 | 57730.40302 | 14.99 | 0.03 | 57730.40359 | 13.95 | 0.04 |
| 57730.40424 | 18.30 | 0.05 | 57730.40487 | 16.03 | 0.04 | 57730.40539 | 15.05 | 0.03 | 57730.40596 | 13.96 | 0.04 |
| 57730.40660 | 18.52 | 0.05 | 57730.40723 | 16.05 | 0.04 | 57730.40777 | 15.00 | 0.03 | 57730.40832 | 13.95 | 0.04 |
| 57730.42765 | 18.24 | 0.05 | 57730.42829 | 16.05 | 0.04 | 57730.42882 | 14.99 | 0.03 | 57730.42937 | 13.96 | 0.04 |
| 57730.43001 | 18.37 | 0.05 | 57730.43065 | 16.06 | 0.04 | 57730.43118 | 15.02 | 0.03 | 57730.43173 | 13.97 | 0.04 |
| 57730.43237 | 18.34 | 0.05 | 57730.43300 | 16.07 | 0.04 | 57730.43353 | 14.98 | 0.03 | 57730.43409 | 13.94 | 0.04 |
| 57730.43472 | 18.21 | 0.05 | 57730.43539 | 16.05 | 0.04 | 57730.43595 | 15.03 | 0.03 | 57730.43650 | 13.93 | 0.04 |
| 57730.43715 | 18.35 | 0.05 | 57730.43782 | 16.09 | 0.04 | 57730.43836 | 14.98 | 0.03 | 57730.43894 | 13.94 | 0.04 |
| 57730.43961 | 18.36 | 0.05 | 57730.44027 | 16.08 | 0.04 | 57730.44084 | 14.97 | 0.03 | 57730.44140 | 13.92 | 0.04 |
| 57730.44206 | 18.33 | 0.05 | 57730.44274 | 16.09 | 0.04 | 57730.44327 | 14.97 | 0.03 | 57730.44384 | 14.00 | 0.04 |
| 57730.44445 | 18.33 | 0.05 | 57730.44511 | 16.03 | 0.04 | 57730.44561 | 15.00 | 0.03 | 57730.44616 | 13.92 | 0.04 |
| 57730.44680 | 18.37 | 0.05 | 57730.44745 | 16.07 | 0.04 | 57730.44797 | 15.06 | 0.03 | 57730.44851 | 13.94 | 0.04 |
| 57730.44915 | 18.29 | 0.05 | 57730.44981 | 16.04 | 0.04 | 57730.45033 | 15.06 | 0.03 | 57730.45090 | 13.92 | 0.04 |
| 57730.45154 | 18.26 | 0.05 | 57730.45218 | 16.10 | 0.04 | 57730.45273 | 15.00 | 0.03 | 57730.45327 | 13.99 | 0.04 |
| 57730.45391 | 18.28 | 0.05 | 57730.45456 | 16.06 | 0.04 | 57730.45510 | 15.00 | 0.03 | 57730.45565 | 13.98 | 0.04 |
| 57730.45629 | 18.34 | 0.05 | 57730.45693 | 16.05 | 0.04 | 57730.45746 | 14.98 | 0.03 | 57730.45801 | 13.97 | 0.04 |
| 57730.45866 | 18.35 | 0.05 | 57730.45930 | 16.14 | 0.04 | 57730.45984 | 15.03 | 0.03 | 57730.46039 | 13.92 | 0.04 |
| 57730.46102 | 18.57 | 0.05 | 57730.46167 | 15.93 | 0.04 | 57730.46220 | 15.02 | 0.03 | 57730.46276 | 13.96 | 0.04 |
| 57730.46339 | 18.39 | 0.05 | 57730.46403 | 16.09 | 0.04 | 57730.46457 | 14.98 | 0.03 | 57730.46512 | 13.98 | 0.04 |
| 57730.46576 | 18.33 | 0.05 | 57730.46638 | 16.11 | 0.04 | 57730.46692 | 15.00 | 0.03 | 57730.46747 | 13.93 | 0.04 |
| 57730.46810 | 18.24 | 0.05 | 57730.46874 | 16.05 | 0.04 | 57730.46927 | 15.00 | 0.03 | 57730.46983 | 13.94 | 0.04 |
| 57730.47047 | 18.32 | 0.05 | 57730.47111 | 16.01 | 0.04 | 57730.47165 | 15.03 | 0.03 | 57730.47221 | 13.91 | 0.04 |
| 57730.47286 | 18.38 | 0.05 | 57730.47350 | 16.09 | 0.04 | 57730.47404 | 15.02 | 0.03 | 57730.47458 | 13.94 | 0.04 |
| 57730.47521 | 18.44 | 0.05 | 57730.47585 | 16.07 | 0.04 | 57730.47639 | 15.00 | 0.03 | 57730.47694 | 13.98 | 0.04 |
| 57730.47757 | 18.42 | 0.05 | 57730.47821 | 16.06 | 0.04 | 57730.47876 | 14.98 | 0.03 | 57730.47931 | 13.97 | 0.04 |
| 57730.47995 | 18.31 | 0.05 | 57730.48059 | 16.10 | 0.04 | 57730.48113 | 15.03 | 0.03 | 57730.48169 | 14.00 | 0.04 |
| 57730.48234 | 18.30 | 0.05 | 57730.48298 | 16.05 | 0.04 | 57730.48353 | 15.04 | 0.03 | 57730.48408 | 13.96 | 0.04 |
| 57730.48472 | 18.47 | 0.05 | 57730.48539 | 16.03 | 0.04 | 57730.48592 | 15.07 | 0.03 | 57730.48648 | 14.00 | 0.04 |
| 57730.48713 | 18.49 | 0.05 | 57730.48779 | 16.07 | 0.04 | 57730.48833 | 15.02 | 0.03 | 57730.48889 | 13.98 | 0.04 |
| 57730.48953 | 18.46 | 0.05 | 57730.49019 | 16.03 | 0.04 | 57730.49073 | 14.99 | 0.03 | 57730.49128 | 13.94 | 0.04 |
| 57730.49192 | 18.47 | 0.05 | 57730.49257 | 16.04 | 0.04 | 57730.49312 | 15.01 | 0.03 | 57730.49367 | 13.94 | 0.04 |
| 57730.49431 | 18.52 | 0.05 | 57730.49497 | 16.04 | 0.04 | 57730.49552 | 15.03 | 0.03 | 57730.49610 | 13.96 | 0.04 |

| Gaia DR1 464387841323087232 ||||||||||||
|---|---|---|---|---|---|---|---|---|---|---|---|
| HJD (2400000 +) | B | Err. (±) | HJD (2400000 +) | V | Err. (±) | HJD (2400000 +) | $R_c$ | Err. (±) | HJD (2400000 +) | $I_c$ | Err. (±) |
| 57730.49674 | 18.29 | 0.04 | 57730.49742 | 16.07 | 0.04 | 57730.49795 | 14.99 | 0.03 | 57730.49853 | 13.97 | 0.04 |
| 57730.49988 | 18.47 | 0.05 | 57730.50055 | 16.00 | 0.04 | 57730.50107 | 15.02 | 0.03 | 57730.50165 | 13.94 | 0.04 |
| 57730.50230 | 18.31 | 0.04 | 57730.50297 | 16.03 | 0.04 | 57730.50350 | 15.03 | 0.03 | 57730.50408 | 13.94 | 0.04 |
| 57730.50473 | 18.49 | 0.05 | 57730.50538 | 16.04 | 0.04 | 57730.50592 | 15.00 | 0.03 | 57730.50650 | 13.94 | 0.04 |
| 57730.50715 | 18.41 | 0.05 | 57730.50784 | 16.04 | 0.04 | 57730.50838 | 15.02 | 0.03 | 57730.50891 | 13.96 | 0.04 |
| 57730.50957 | 18.46 | 0.05 | 57730.51022 | 16.04 | 0.04 | 57730.51076 | 14.97 | 0.03 | 57730.51135 | 13.95 | 0.04 |
| 57730.51199 | 18.40 | 0.05 | 57730.51265 | 16.00 | 0.04 | 57730.51319 | 15.02 | 0.03 | 57730.51375 | 13.94 | 0.04 |
| 57730.51442 | 18.43 | 0.05 | 57730.51507 | 16.07 | 0.04 | 57730.51563 | 15.02 | 0.03 | 57730.51618 | 13.94 | 0.04 |
| 57730.51684 | 18.34 | 0.05 | 57730.51750 | 16.06 | 0.04 | 57730.51804 | 14.99 | 0.03 | 57730.51861 | 13.95 | 0.04 |
| 57730.51928 | 18.38 | 0.05 | 57730.51992 | 16.08 | 0.04 | 57730.52047 | 15.00 | 0.03 | 57730.52104 | 14.01 | 0.04 |
| 57730.52172 | 18.29 | 0.05 | 57730.52239 | 16.08 | 0.04 | 57730.52294 | 15.01 | 0.03 | 57730.52351 | 13.96 | 0.04 |
| 57730.52419 | 18.43 | 0.05 | 57730.52484 | 16.05 | 0.04 | 57730.52539 | 14.99 | 0.03 | 57730.52599 | 13.97 | 0.04 |
| 57730.52665 | 18.47 | 0.05 | 57730.52732 | 16.08 | 0.04 | | | | | | |
| 57730.52877 | 18.27 | 0.05 | 57730.52937 | 16.08 | 0.04 | 57730.52986 | 14.98 | 0.03 | 57730.53037 | 13.89 | 0.04 |
| 57730.53097 | 18.40 | 0.05 | 57730.53158 | 16.07 | 0.04 | 57730.53207 | 15.01 | 0.03 | 57730.53258 | 13.96 | 0.04 |
| 57730.53317 | 18.36 | 0.05 | 57730.53379 | 16.03 | 0.04 | 57730.53428 | 14.99 | 0.03 | 57730.53479 | 13.93 | 0.04 |
| 57730.53540 | 18.42 | 0.05 | 57730.53601 | 16.09 | 0.04 | 57730.53650 | 15.08 | 0.03 | 57730.53702 | 13.92 | 0.04 |
| 57730.53762 | 18.29 | 0.05 | 57730.53823 | 16.07 | 0.04 | 57730.53872 | 15.03 | 0.03 | 57730.53925 | 13.96 | 0.04 |
| 57730.53985 | 18.48 | 0.05 | 57730.54045 | 15.99 | 0.04 | 57730.54096 | 15.09 | 0.03 | 57730.54147 | 13.96 | 0.04 |
| 57730.54208 | 18.39 | 0.05 | 57730.54268 | 16.08 | 0.04 | 57730.54318 | 14.98 | 0.03 | 57730.54371 | 13.99 | 0.04 |
| 57730.54431 | 18.34 | 0.05 | 57730.54492 | 16.10 | 0.04 | 57730.54542 | 15.04 | 0.03 | 57730.54594 | 13.91 | 0.04 |
| 57730.54656 | 18.32 | 0.05 | 57730.54717 | 16.04 | 0.04 | 57730.54767 | 15.03 | 0.03 | 57730.54820 | 13.95 | 0.04 |
| 57730.54882 | 18.40 | 0.05 | 57730.54943 | 16.02 | 0.04 | 57730.54994 | 15.05 | 0.03 | 57730.55046 | 13.93 | 0.04 |
| 57730.55107 | 18.79 | 0.06 | 57730.55171 | 15.95 | 0.04 | 57730.55221 | 15.06 | 0.03 | 57730.55274 | 13.98 | 0.04 |
| 57730.55336 | 18.48 | 0.06 | 57730.55398 | 16.00 | 0.04 | 57730.55449 | 15.00 | 0.03 | 57730.55502 | 13.94 | 0.04 |
| 57730.55564 | 18.26 | 0.05 | 57730.55626 | 16.10 | 0.04 | 57730.55678 | 14.97 | 0.03 | 57730.55730 | 13.99 | 0.04 |
| 57730.55794 | 18.31 | 0.05 | 57730.55855 | 16.09 | 0.04 | 57730.55907 | 15.04 | 0.03 | 57730.55961 | 13.94 | 0.04 |
| 57730.56022 | 18.33 | 0.05 | 57730.56085 | 16.07 | 0.04 | 57730.56138 | 15.03 | 0.03 | 57730.56190 | 13.93 | 0.04 |
| 57730.56252 | 18.61 | 0.05 | 57730.56316 | 16.10 | 0.04 | 57730.56366 | 14.94 | 0.03 | 57730.56420 | 14.01 | 0.04 |
| 57730.56483 | 18.33 | 0.05 | 57730.56546 | 15.95 | 0.04 | 57730.56599 | 15.06 | 0.03 | 57730.56653 | 13.97 | 0.04 |
| 57730.56716 | 18.46 | 0.05 | 57730.56778 | 16.08 | 0.04 | 57730.56832 | 15.02 | 0.03 | 57730.56884 | 13.91 | 0.04 |
| 57730.56947 | 18.40 | 0.05 | 57730.57012 | 16.05 | 0.04 | 57730.57063 | 14.97 | 0.03 | 57730.57117 | 13.94 | 0.04 |
| 57730.57180 | 18.61 | 0.05 | 57730.57244 | 16.06 | 0.04 | 57730.57297 | 14.99 | 0.03 | 57730.57352 | 13.95 | 0.04 |
| 57730.57415 | 18.39 | 0.05 | 57730.57479 | 16.11 | 0.04 | 57730.57531 | 15.14 | 0.03 | 57730.57586 | 13.96 | 0.04 |
| 57730.57649 | 18.25 | 0.05 | 57730.57712 | 16.03 | 0.04 | 57730.57765 | 15.03 | 0.03 | 57730.57818 | 13.97 | 0.04 |
| 57730.57881 | 18.69 | 0.05 | 57730.57946 | 16.04 | 0.04 | 57730.57998 | 15.03 | 0.03 | 57730.58052 | 13.96 | 0.04 |
| 57730.58115 | 18.52 | 0.05 | 57730.58180 | 16.10 | 0.04 | 57730.58232 | 14.93 | 0.03 | 57730.58287 | 13.92 | 0.04 |
| 57730.58349 | 18.52 | 0.05 | 57730.58414 | 16.12 | 0.04 | 57730.58467 | 15.02 | 0.03 | 57730.58522 | 13.92 | 0.04 |
| 57730.58585 | 18.43 | 0.05 | 57730.58648 | 16.06 | 0.04 | 57730.58702 | 15.00 | 0.03 | 57730.58756 | 13.94 | 0.04 |
| 57730.58818 | 18.32 | 0.05 | 57730.58882 | 16.03 | 0.04 | 57730.58935 | 14.92 | 0.03 | 57730.58990 | 13.98 | 0.04 |
| 57730.59052 | 18.32 | 0.05 | 57730.59118 | 16.11 | 0.04 | 57730.59171 | 14.97 | 0.03 | 57730.59225 | 13.95 | 0.04 |
| 57730.59289 | 18.39 | 0.05 | 57730.59352 | 16.15 | 0.04 | 57730.59405 | 15.03 | 0.03 | 57730.59458 | 13.94 | 0.04 |
| 57730.59521 | 18.33 | 0.05 | 57730.59586 | 16.09 | 0.04 | 57730.59637 | 14.96 | 0.03 | 57730.59692 | 14.05 | 0.04 |
| 57730.59778 | 18.44 | 0.05 | 57730.59843 | 16.10 | 0.04 | 57730.59894 | 15.02 | 0.03 | 57730.59949 | 13.95 | 0.04 |
| 57730.60013 | 18.42 | 0.05 | 57730.60078 | 16.05 | 0.04 | 57730.60129 | 15.05 | 0.03 | 57730.60184 | 13.91 | 0.04 |
| 57730.60248 | 18.44 | 0.05 | 57730.60310 | 16.04 | 0.04 | 57730.60365 | 15.06 | 0.03 | 57730.60419 | 13.89 | 0.04 |
| 57730.60483 | 18.55 | 0.05 | 57730.60554 | 16.03 | 0.04 | 57730.60608 | 15.04 | 0.03 | 57730.60666 | 13.90 | 0.04 |
| 57730.60731 | 18.45 | 0.05 | 57730.60800 | 15.97 | 0.04 | 57730.60855 | 15.05 | 0.03 | 57730.60910 | 14.00 | 0.04 |
| 57730.60973 | 18.36 | 0.05 | 57730.61037 | 16.08 | 0.04 | 57730.61091 | 15.02 | 0.03 | 57730.61146 | 13.82 | 0.04 |
| 57730.61211 | 18.41 | 0.05 | 57730.61277 | 16.01 | 0.04 | 57730.61330 | 15.01 | 0.03 | 57730.61385 | 13.92 | 0.04 |
| 57730.61462 | 18.39 | 0.05 | 57730.61525 | 16.04 | 0.04 | 57730.61579 | 15.02 | 0.03 | 57730.61634 | 13.95 | 0.04 |
| 57730.61697 | 18.48 | 0.05 | 57730.61763 | 16.13 | 0.04 | 57730.61817 | 14.96 | 0.03 | 57730.61873 | 14.00 | 0.04 |
| 57730.61937 | 18.47 | 0.05 | 57730.62000 | 16.06 | 0.04 | 57730.62054 | 15.04 | 0.03 | 57730.62110 | 13.91 | 0.04 |
| 57730.62174 | 18.25 | 0.05 | 57730.62238 | 16.05 | 0.04 | 57730.62292 | 15.03 | 0.03 | 57730.62346 | 13.97 | 0.04 |
| 57730.62410 | 18.41 | 0.05 | 57730.62475 | 16.06 | 0.04 | 57730.62528 | 15.02 | 0.03 | 57730.62583 | 13.93 | 0.04 |
| 57730.62646 | 18.32 | 0.05 | 57730.62712 | 16.12 | 0.04 | 57730.62764 | 15.06 | 0.03 | 57730.62819 | 13.99 | 0.04 |
| 57730.62883 | 18.42 | 0.05 | 57730.62947 | 15.98 | 0.04 | 57730.63001 | 15.09 | 0.03 | 57730.63056 | 13.96 | 0.04 |
| 57730.63119 | 18.37 | 0.05 | 57730.63182 | 16.14 | 0.04 | 57730.63236 | 14.92 | 0.03 | 57730.63289 | 13.96 | 0.04 |
| 57730.63352 | 18.17 | 0.05 | 57730.63417 | 16.16 | 0.04 | 57730.63470 | 15.03 | 0.03 | 57730.63525 | 13.96 | 0.04 |
| 57730.63589 | 18.39 | 0.05 | 57730.63654 | 16.16 | 0.04 | 57730.63707 | 15.00 | 0.03 | 57730.63763 | 13.91 | 0.04 |
| 57730.63855 | 18.46 | 0.05 | 57730.63914 | 16.02 | 0.04 | 57730.63964 | 15.02 | 0.03 | 57730.64015 | 13.98 | 0.04 |
| 57730.64075 | 18.41 | 0.05 | 57730.64134 | 16.10 | 0.04 | 57730.64183 | 15.02 | 0.03 | 57730.64235 | 13.87 | 0.04 |
| 57730.64296 | 18.56 | 0.06 | 57730.64356 | 15.95 | 0.04 | 57730.64406 | 15.00 | 0.03 | 57730.64457 | 13.95 | 0.04 |
| 57730.64540 | 18.62 | 0.05 | 57730.64601 | 16.01 | 0.04 | 57730.64651 | 14.99 | 0.03 | 57730.64703 | 13.96 | 0.04 |
| 57730.64801 | 18.52 | 0.05 | 57730.64862 | 16.16 | 0.04 | 57730.64911 | 15.05 | 0.03 | 57730.64964 | 13.99 | 0.04 |

| Gaia DR1 4643878413230872322 ||||||||
|---|---|---|---|---|---|---|---|---|
| HJD (2400000 +) | B-V | Err. (±) | HJD (2400000 +) | V-R | Err. (±) | HJD (2400000 +) | V-I | Err. (±) |
| 57730.34008 | 2.21 | 0.07 | 57730.34063 | 1.02 | 0.05 | 57730.34089 | 2.05 | 0.06 |
| 57730.34230 | 2.45 | 0.07 | 57730.34285 | 1.03 | 0.05 | 57730.34311 | 2.10 | 0.06 |
| 57730.34452 | 2.36 | 0.07 | 57730.34507 | 1.06 | 0.05 | 57730.34533 | 2.13 | 0.06 |
| 57730.34676 | 2.32 | 0.07 | 57730.34732 | 1.01 | 0.05 | 57730.34758 | 2.03 | 0.06 |
| 57730.34901 | 2.14 | 0.07 | 57730.34956 | 1.10 | 0.05 | 57730.34983 | 2.13 | 0.06 |
| 57730.35126 | 2.24 | 0.07 | 57730.35182 | 1.11 | 0.05 | 57730.35209 | 2.15 | 0.06 |
| 57730.35352 | 2.29 | 0.07 | 57730.35408 | 1.07 | 0.05 | 57730.35434 | 2.13 | 0.06 |
| 57730.35578 | 2.22 | 0.07 | 57730.35635 | 1.06 | 0.05 | 57730.35661 | 2.12 | 0.06 |
| 57730.35806 | 2.86 | 0.07 | 57730.35863 | 0.68 | 0.05 | 57730.35889 | 1.78 | 0.06 |
| 57730.36034 | 2.20 | 0.06 | 57730.36090 | 1.05 | 0.05 | 57730.36117 | 2.12 | 0.06 |
| 57730.36262 | 2.29 | 0.07 | 57730.36266 | 1.06 | 0.05 | 57730.36346 | 2.13 | 0.06 |
| 57730.36492 | 2.43 | 0.07 | 57730.36549 | 0.99 | 0.05 | 57730.36576 | 2.08 | 0.06 |
| 57730.36721 | 2.46 | 0.07 | 57730.36778 | 1.00 | 0.05 | 57730.36805 | 2.09 | 0.06 |
| 57730.36951 | 2.29 | 0.07 | 57730.37008 | 1.07 | 0.05 | 57730.37035 | 2.08 | 0.06 |
| 57730.37183 | 2.35 | 0.07 | 57730.37240 | 1.03 | 0.05 | 57730.37268 | 2.13 | 0.06 |
| 57730.37415 | 2.40 | 0.07 | 57730.37472 | 1.04 | 0.05 | 57730.37499 | 2.13 | 0.06 |
| 57730.37647 | 2.38 | 0.07 | 57730.37705 | 1.05 | 0.05 | 57730.37731 | 2.16 | 0.06 |
| 57730.37879 | 2.47 | 0.07 | 57730.37936 | 1.04 | 0.05 | 57730.37963 | 2.14 | 0.06 |
| 57730.38110 | 2.26 | 0.06 | 57730.38167 | 1.08 | 0.05 | 57730.38194 | 2.10 | 0.06 |
| 57730.38341 | 2.30 | 0.07 | 57730.38400 | 1.07 | 0.05 | 57730.38427 | 2.10 | 0.06 |
| 57730.38576 | 2.20 | 0.07 | 57730.38635 | 1.05 | 0.05 | 57730.38662 | 2.08 | 0.06 |
| 57730.38812 | 2.37 | 0.07 | 57730.38869 | 0.99 | 0.05 | 57730.38897 | 2.05 | 0.06 |
| 57730.39047 | 2.34 | 0.06 | 57730.39105 | 0.97 | 0.05 | 57730.39132 | 2.09 | 0.06 |
| 57730.39281 | 2.43 | 0.07 | 57730.39339 | 1.02 | 0.05 | 57730.39367 | 2.06 | 0.06 |
| 57730.39514 | 2.14 | 0.06 | 57730.39572 | 1.03 | 0.05 | 57730.39599 | 2.13 | 0.06 |
| 57730.39747 | 2.36 | 0.06 | 57730.39805 | 1.07 | 0.05 | 57730.39832 | 2.08 | 0.06 |
| 57730.39981 | 2.41 | 0.07 | 57730.40040 | 1.10 | 0.05 | 57730.40067 | 2.10 | 0.06 |
| 57730.40216 | 2.55 | 0.07 | 57730.40275 | 1.03 | 0.05 | 57730.40303 | 2.06 | 0.06 |
| 57730.40455 | 2.26 | 0.06 | 57730.40513 | 0.99 | 0.05 | 57730.40541 | 2.08 | 0.06 |
| 57730.40691 | 2.47 | 0.07 | 57730.40750 | 1.06 | 0.05 | 57730.40777 | 2.11 | 0.06 |
| 57730.42797 | 2.20 | 0.07 | 57730.42855 | 1.05 | 0.05 | 57730.42883 | 2.08 | 0.06 |
| 57730.43033 | 2.31 | 0.06 | 57730.43092 | 1.05 | 0.05 | 57730.43119 | 2.09 | 0.06 |
| 57730.43269 | 2.27 | 0.07 | 57730.43327 | 1.09 | 0.05 | 57730.43355 | 2.13 | 0.06 |
| 57730.43505 | 2.16 | 0.07 | 57730.43567 | 1.02 | 0.05 | 57730.43594 | 2.12 | 0.06 |
| 57730.43748 | 2.26 | 0.07 | 57730.43809 | 1.10 | 0.05 | 57730.43838 | 2.15 | 0.06 |
| 57730.43994 | 2.29 | 0.07 | 57730.44056 | 1.11 | 0.05 | 57730.44084 | 2.16 | 0.06 |
| 57730.44240 | 2.24 | 0.07 | 57730.44300 | 1.11 | 0.05 | 57730.44329 | 2.09 | 0.06 |
| 57730.44478 | 2.29 | 0.07 | 57730.44536 | 1.04 | 0.05 | 57730.44563 | 2.11 | 0.06 |
| 57730.44713 | 2.29 | 0.07 | 57730.44771 | 1.02 | 0.05 | 57730.44798 | 2.13 | 0.06 |
| 57730.44948 | 2.25 | 0.06 | 57730.45007 | 0.99 | 0.05 | 57730.45035 | 2.12 | 0.06 |
| 57730.45186 | 2.16 | 0.06 | 57730.45245 | 1.09 | 0.05 | 57730.45272 | 2.11 | 0.06 |
| 57730.45424 | 2.22 | 0.06 | 57730.45483 | 1.06 | 0.05 | 57730.45511 | 2.09 | 0.06 |
| 57730.45661 | 2.29 | 0.06 | 57730.45719 | 1.07 | 0.05 | 57730.45747 | 2.08 | 0.06 |
| 57730.45898 | 2.21 | 0.07 | 57730.45957 | 1.10 | 0.05 | 57730.45984 | 2.21 | 0.06 |
| 57730.46134 | 2.63 | 0.07 | 57730.46193 | 0.94 | 0.05 | 57730.46221 | 1.99 | 0.06 |
| 57730.46371 | 2.29 | 0.07 | 57730.46430 | 1.11 | 0.05 | 57730.46458 | 2.12 | 0.06 |
| 57730.46607 | 2.23 | 0.06 | 57730.46665 | 1.10 | 0.05 | 57730.46693 | 2.18 | 0.06 |
| 57730.46842 | 2.19 | 0.06 | 57730.46901 | 1.04 | 0.05 | 57730.46929 | 2.10 | 0.06 |
| 57730.47079 | 2.31 | 0.07 | 57730.47138 | 0.99 | 0.05 | 57730.47166 | 2.10 | 0.06 |
| 57730.47318 | 2.29 | 0.06 | 57730.47377 | 1.08 | 0.05 | 57730.47404 | 2.15 | 0.06 |
| 57730.47553 | 2.38 | 0.07 | 57730.47612 | 1.07 | 0.05 | 57730.47639 | 2.09 | 0.06 |
| 57730.47789 | 2.36 | 0.06 | 57730.47848 | 1.08 | 0.05 | 57730.47876 | 2.09 | 0.06 |
| 57730.48027 | 2.22 | 0.06 | 57730.48086 | 1.06 | 0.05 | 57730.48114 | 2.09 | 0.06 |
| 57730.48266 | 2.24 | 0.06 | 57730.48325 | 1.02 | 0.05 | 57730.48353 | 2.09 | 0.06 |
| 57730.48505 | 2.44 | 0.06 | 57730.48565 | 0.97 | 0.05 | 57730.48594 | 2.03 | 0.06 |
| 57730.48746 | 2.43 | 0.07 | 57730.48806 | 1.06 | 0.05 | 57730.48834 | 2.09 | 0.06 |
| 57730.48986 | 2.43 | 0.06 | 57730.49046 | 1.05 | 0.05 | 57730.49073 | 2.09 | 0.06 |
| 57730.49225 | 2.44 | 0.06 | 57730.49285 | 1.04 | 0.05 | 57730.49312 | 2.10 | 0.06 |
| 57730.49464 | 2.48 | 0.06 | 57730.49524 | 1.02 | 0.05 | 57730.49554 | 2.08 | 0.06 |
| 57730.49708 | 2.23 | 0.06 | 57730.49769 | 1.07 | 0.05 | 57730.49798 | 2.10 | 0.06 |
| 57730.50021 | 2.47 | 0.06 | 57730.50081 | 0.99 | 0.05 | 57730.50110 | 2.07 | 0.06 |
| 57730.50263 | 2.28 | 0.06 | 57730.50323 | 1.01 | 0.05 | 57730.50352 | 2.10 | 0.06 |
| 57730.50506 | 2.45 | 0.06 | 57730.50565 | 1.05 | 0.05 | 57730.50594 | 2.11 | 0.06 |
| 57730.50749 | 2.37 | 0.06 | 57730.50811 | 1.03 | 0.05 | 57730.50838 | 2.09 | 0.06 |

| Gaia DR1 464387841323087232 ||||||||
| HJD (2400000 +) | B-V | Err. (±) | HJD (2400000 +) | V-R | Err. (±) | HJD (2400000 +) | V-I | Err. (±) |
|---|---|---|---|---|---|---|---|---|
| 57730.50990 | 2.43 | 0.06 | 57730.51049 | 1.08 | 0.05 | 57730.51078 | 2.09 | 0.06 |
| 57730.51232 | 2.39 | 0.06 | 57730.51292 | 0.99 | 0.05 | 57730.51320 | 2.07 | 0.06 |
| 57730.51475 | 2.36 | 0.06 | 57730.51535 | 1.05 | 0.05 | 57730.51563 | 2.13 | 0.06 |
| 57730.51717 | 2.28 | 0.06 | 57730.51777 | 1.07 | 0.05 | 57730.51806 | 2.11 | 0.06 |
| 57730.51960 | 2.30 | 0.06 | 57730.52020 | 1.08 | 0.05 | 57730.52048 | 2.07 | 0.06 |
| 57730.52206 | 2.20 | 0.06 | 57730.52266 | 1.07 | 0.05 | 57730.52295 | 2.12 | 0.06 |
| 57730.52451 | 2.38 | 0.07 | 57730.52512 | 1.06 | 0.05 | 57730.52541 | 2.09 | 0.06 |
| 57730.52698 | 2.39 | 0.06 | | | | | | |
| 57730.52907 | 2.19 | 0.06 | 57730.52961 | 1.09 | 0.05 | 57730.52987 | 2.19 | 0.06 |
| 57730.53127 | 2.33 | 0.06 | 57730.53182 | 1.06 | 0.05 | 57730.53208 | 2.11 | 0.06 |
| 57730.53348 | 2.33 | 0.06 | 57730.53403 | 1.04 | 0.05 | 57730.53429 | 2.10 | 0.06 |
| 57730.53571 | 2.32 | 0.06 | 57730.53626 | 1.02 | 0.05 | 57730.53651 | 2.17 | 0.06 |
| 57730.53793 | 2.22 | 0.06 | 57730.53848 | 1.04 | 0.05 | 57730.53874 | 2.11 | 0.06 |
| 57730.54015 | 2.49 | 0.06 | 57730.54071 | 0.92 | 0.05 | 57730.54096 | 2.04 | 0.06 |
| 57730.54238 | 2.31 | 0.06 | 57730.54293 | 1.10 | 0.06 | 57730.54319 | 2.09 | 0.06 |
| 57730.54462 | 2.24 | 0.06 | 57730.54517 | 1.06 | 0.05 | 57730.54543 | 2.18 | 0.06 |
| 57730.54686 | 2.28 | 0.06 | 57730.54742 | 1.02 | 0.05 | 57730.54768 | 2.09 | 0.06 |
| 57730.54913 | 2.38 | 0.07 | 57730.54969 | 0.98 | 0.05 | 57730.54995 | 2.09 | 0.06 |
| 57730.55139 | 2.84 | 0.07 | 57730.55196 | 0.93 | 0.05 | 57730.55223 | 2.00 | 0.06 |
| 57730.55367 | 2.48 | 0.07 | 57730.55423 | 1.02 | 0.05 | 57730.55450 | 2.07 | 0.06 |
| 57730.55595 | 2.16 | 0.06 | 57730.55652 | 1.13 | 0.05 | 57730.55678 | 2.10 | 0.06 |
| 57730.55825 | 2.21 | 0.06 | 57730.55881 | 1.06 | 0.05 | 57730.55908 | 2.15 | 0.06 |
| 57730.56053 | 2.26 | 0.06 | 57730.56112 | 1.04 | 0.05 | 57730.56137 | 2.14 | 0.06 |
| 57730.56284 | 2.52 | 0.07 | 57730.56341 | 1.16 | 0.05 | 57730.56368 | 2.10 | 0.06 |
| 57730.56515 | 2.38 | 0.06 | 57730.56573 | 0.90 | 0.05 | 57730.56600 | 1.98 | 0.06 |
| 57730.56747 | 2.39 | 0.06 | 57730.56805 | 1.07 | 0.05 | 57730.56831 | 2.17 | 0.06 |
| 57730.56980 | 2.34 | 0.06 | 57730.57037 | 1.08 | 0.05 | 57730.57064 | 2.11 | 0.06 |
| 57730.57212 | 2.55 | 0.07 | 57730.57271 | 1.08 | 0.05 | 57730.57298 | 2.13 | 0.06 |
| 57730.57447 | 2.28 | 0.06 | 57730.57505 | 0.98 | 0.05 | 57730.57533 | 2.15 | 0.06 |
| 57730.57680 | 2.22 | 0.06 | 57730.57738 | 1.00 | 0.05 | 57730.57765 | 2.06 | 0.06 |
| 57730.57914 | 2.65 | 0.07 | 57730.57972 | 1.03 | 0.05 | 57730.57999 | 2.09 | 0.06 |
| 57730.58147 | 2.42 | 0.07 | 57730.58206 | 1.17 | 0.05 | 57730.58233 | 2.19 | 0.06 |
| 57730.58382 | 2.39 | 0.07 | 57730.58441 | 1.11 | 0.05 | 57730.58468 | 2.20 | 0.06 |
| 57730.58617 | 2.37 | 0.07 | 57730.58675 | 1.06 | 0.05 | 57730.58702 | 2.12 | 0.06 |
| 57730.58850 | 2.29 | 0.06 | 57730.58908 | 1.11 | 0.05 | 57730.58936 | 2.05 | 0.06 |
| 57730.59085 | 2.21 | 0.06 | 57730.59144 | 1.13 | 0.05 | 57730.59171 | 2.16 | 0.06 |
| 57730.59320 | 2.24 | 0.07 | 57730.59378 | 1.12 | 0.05 | 57730.59405 | 2.20 | 0.06 |
| 57730.59554 | 2.24 | 0.06 | 57730.59611 | 1.13 | 0.05 | 57730.59639 | 2.04 | 0.06 |
| 57730.59811 | 2.34 | 0.06 | 57730.59868 | 1.08 | 0.05 | 57730.59896 | 2.15 | 0.06 |
| 57730.60045 | 2.37 | 0.07 | 57730.60103 | 1.01 | 0.05 | 57730.60131 | 2.14 | 0.06 |
| 57730.60279 | 2.41 | 0.07 | 57730.60338 | 0.98 | 0.05 | 57730.60365 | 2.15 | 0.06 |
| 57730.60519 | 2.52 | 0.07 | 57730.60581 | 1.00 | 0.06 | 57730.60610 | 2.14 | 0.06 |
| 57730.60765 | 2.48 | 0.07 | 57730.60827 | 0.93 | 0.05 | 57730.60855 | 1.98 | 0.06 |
| 57730.61005 | 2.28 | 0.07 | 57730.61064 | 1.06 | 0.05 | 57730.61091 | 2.26 | 0.06 |
| 57730.61244 | 2.40 | 0.07 | 57730.61304 | 1.01 | 0.05 | 57730.61331 | 2.09 | 0.06 |
| 57730.61494 | 2.35 | 0.06 | 57730.61552 | 1.02 | 0.05 | 57730.61580 | 2.09 | 0.06 |
| 57730.61730 | 2.36 | 0.07 | 57730.61790 | 1.16 | 0.06 | 57730.61818 | 2.13 | 0.06 |
| 57730.61968 | 2.41 | 0.07 | 57730.62027 | 1.03 | 0.05 | 57730.62055 | 2.15 | 0.06 |
| 57730.62206 | 2.21 | 0.07 | 57730.62265 | 1.02 | 0.05 | 57730.62292 | 2.07 | 0.06 |
| 57730.62442 | 2.36 | 0.07 | 57730.62501 | 1.04 | 0.05 | 57730.62529 | 2.12 | 0.06 |
| 57730.62679 | 2.19 | 0.07 | 57730.62738 | 1.06 | 0.06 | 57730.62766 | 2.13 | 0.06 |
| 57730.62915 | 2.44 | 0.07 | 57730.62974 | 0.91 | 0.05 | 57730.63001 | 2.03 | 0.06 |
| 57730.63150 | 2.23 | 0.07 | 57730.63209 | 1.20 | 0.06 | 57730.63236 | 2.17 | 0.06 |
| 57730.63385 | 2.02 | 0.07 | 57730.63444 | 1.11 | 0.06 | 57730.63471 | 2.18 | 0.06 |
| 57730.63621 | 2.23 | 0.07 | 57730.63680 | 1.15 | 0.06 | 57730.63708 | 2.24 | 0.06 |
| 57730.63885 | 2.44 | 0.07 | 57730.63939 | 1.01 | 0.05 | 57730.63965 | 2.05 | 0.06 |
| 57730.64104 | 2.32 | 0.07 | 57730.64159 | 1.08 | 0.05 | 57730.64185 | 2.22 | 0.06 |
| 57730.64326 | 2.61 | 0.07 | 57730.64381 | 0.96 | 0.05 | 57730.64407 | 2.01 | 0.06 |
| 57730.64570 | 2.60 | 0.07 | 57730.64626 | 1.04 | 0.05 | 57730.64652 | 2.07 | 0.06 |
| 57730.64832 | 2.35 | 0.07 | 57730.64887 | 1.11 | 0.06 | 57730.64913 | 2.18 | 0.06 |

# Appendix 3
# BVR$_c$I$_c$ photometric data of star Gaia DR1 464387841322326272

| Gaia DR1 464387841322326272 | | | | | | | | | | |
|---|---|---|---|---|---|---|---|---|---|---|
| HJD (2400000 +) | B | Err. (±) | HJD (2400000 +) | V | Err. (±) | HJD (2400000 +) | R$_c$ | Err. (±) | HJD (2400000 +) | I$_c$ | Err. (±) |
| 57730.33978 | 17.08 | 0.04 | 57730.34039 | 15.77 | 0.04 | 57730.34088 | 15.29 | 0.03 | 57730.34139 | 14.64 | 0.04 |
| 57730.34199 | 17.14 | 0.04 | 57730.34260 | 15.77 | 0.04 | 57730.34309 | 15.28 | 0.03 | 57730.34361 | 14.65 | 0.04 |
| 57730.34421 | 17.21 | 0.04 | 57730.34482 | 15.84 | 0.04 | 57730.34532 | 15.30 | 0.03 | 57730.34585 | 14.65 | 0.04 |
| 57730.34646 | 17.16 | 0.04 | 57730.34707 | 15.83 | 0.04 | 57730.34757 | 15.26 | 0.03 | 57730.34809 | 14.66 | 0.04 |
| 57730.34870 | 17.06 | 0.04 | 57730.34931 | 15.83 | 0.04 | 57730.34982 | 15.23 | 0.03 | 57730.35035 | 14.63 | 0.04 |
| 57730.35095 | 17.09 | 0.04 | 57730.35157 | 15.85 | 0.04 | 57730.35207 | 15.25 | 0.03 | 57730.35260 | 14.63 | 0.04 |
| 57730.35322 | 17.16 | 0.04 | 57730.35383 | 15.78 | 0.04 | 57730.35434 | 15.21 | 0.03 | 57730.35486 | 14.65 | 0.04 |
| 57730.35547 | 17.08 | 0.04 | 57730.35609 | 15.78 | 0.04 | 57730.35660 | 15.33 | 0.03 | 57730.35713 | 14.63 | 0.04 |
| 57730.35776 | 17.31 | 0.04 | 57730.35837 | 15.79 | 0.04 | 57730.35888 | 15.31 | 0.03 | 57730.35941 | 14.65 | 0.04 |
| 57730.36002 | 17.22 | 0.04 | 57730.36065 | 15.78 | 0.04 | 57730.36115 | 15.28 | 0.03 | 57730.36169 | 14.66 | 0.04 |
| 57730.36232 | 17.24 | 0.04 | 57730.36293 | 15.80 | 0.04 | 57730.36239 | 15.28 | 0.03 | 57730.36398 | 14.67 | 0.04 |
| 57730.36461 | 17.25 | 0.04 | 57730.36524 | 15.81 | 0.04 | 57730.36574 | 15.23 | 0.03 | 57730.36627 | 14.64 | 0.04 |
| 57730.36690 | 17.16 | 0.04 | 57730.36752 | 15.79 | 0.04 | 57730.36804 | 15.30 | 0.03 | 57730.36858 | 14.59 | 0.04 |
| 57730.36920 | 17.20 | 0.04 | 57730.36982 | 15.77 | 0.04 | 57730.37034 | 15.25 | 0.03 | 57730.37088 | 14.61 | 0.04 |
| 57730.37152 | 17.08 | 0.04 | 57730.37215 | 15.83 | 0.04 | 57730.37265 | 15.25 | 0.03 | 57730.37321 | 14.61 | 0.04 |
| 57730.37383 | 17.21 | 0.04 | 57730.37446 | 15.83 | 0.04 | 57730.37498 | 15.25 | 0.03 | 57730.37552 | 14.63 | 0.04 |
| 57730.37615 | 17.15 | 0.04 | 57730.37679 | 15.81 | 0.04 | 57730.37730 | 15.32 | 0.03 | 57730.37784 | 14.64 | 0.04 |
| 57730.37847 | 17.24 | 0.04 | 57730.37910 | 15.77 | 0.04 | 57730.37962 | 15.29 | 0.03 | 57730.38015 | 14.61 | 0.04 |
| 57730.38079 | 17.17 | 0.04 | 57730.38141 | 15.80 | 0.04 | 57730.38193 | 15.26 | 0.03 | 57730.38248 | 14.68 | 0.04 |
| 57730.38309 | 17.19 | 0.04 | 57730.38373 | 15.81 | 0.04 | 57730.38426 | 15.21 | 0.03 | 57730.38481 | 14.65 | 0.04 |
| 57730.38544 | 17.06 | 0.04 | 57730.38608 | 15.80 | 0.04 | 57730.38661 | 15.27 | 0.03 | 57730.38716 | 14.65 | 0.04 |
| 57730.38780 | 17.12 | 0.04 | 57730.38843 | 15.77 | 0.04 | 57730.38896 | 15.30 | 0.03 | 57730.38951 | 14.64 | 0.04 |
| 57730.39015 | 17.26 | 0.04 | 57730.39079 | 15.79 | 0.04 | 57730.39130 | 15.32 | 0.03 | 57730.39185 | 14.65 | 0.04 |
| 57730.39249 | 17.15 | 0.04 | 57730.39313 | 15.80 | 0.04 | 57730.39366 | 15.30 | 0.03 | 57730.39420 | 14.65 | 0.04 |
| 57730.39483 | 17.06 | 0.04 | 57730.39546 | 15.79 | 0.04 | 57730.39598 | 15.30 | 0.03 | 57730.39653 | 14.61 | 0.04 |
| 57730.39715 | 17.27 | 0.04 | 57730.39778 | 15.79 | 0.04 | 57730.39831 | 15.25 | 0.03 | 57730.39885 | 14.64 | 0.04 |
| 57730.39949 | 17.14 | 0.04 | 57730.40013 | 15.84 | 0.04 | 57730.40066 | 15.26 | 0.03 | 57730.40120 | 14.65 | 0.04 |
| 57730.40184 | 17.15 | 0.04 | 57730.40248 | 15.78 | 0.04 | 57730.40302 | 15.28 | 0.03 | 57730.40359 | 14.63 | 0.04 |
| 57730.40424 | 17.25 | 0.04 | 57730.40487 | 15.79 | 0.04 | 57730.40539 | 15.29 | 0.03 | 57730.40596 | 14.65 | 0.04 |
| 57730.40660 | 17.14 | 0.04 | 57730.40723 | 15.81 | 0.04 | 57730.40777 | 15.27 | 0.03 | 57730.40832 | 14.67 | 0.04 |
| 57730.42765 | 17.00 | 0.04 | 57730.42829 | 15.81 | 0.04 | 57730.42882 | 15.27 | 0.03 | 57730.42937 | 14.61 | 0.04 |
| 57730.43001 | 17.09 | 0.04 | 57730.43065 | 15.80 | 0.04 | 57730.43118 | 15.27 | 0.03 | 57730.43173 | 14.66 | 0.04 |
| 57730.43237 | 17.18 | 0.04 | 57730.43300 | 15.83 | 0.04 | 57730.43353 | 15.21 | 0.03 | 57730.43409 | 14.64 | 0.04 |
| 57730.43472 | 17.07 | 0.04 | 57730.43539 | 15.80 | 0.04 | 57730.43595 | 15.34 | 0.03 | 57730.43650 | 14.66 | 0.04 |
| 57730.43715 | 17.11 | 0.04 | 57730.43782 | 15.82 | 0.04 | 57730.43836 | 15.24 | 0.03 | 57730.43894 | 14.61 | 0.04 |
| 57730.43961 | 17.04 | 0.04 | 57730.44027 | 15.81 | 0.04 | 57730.44084 | 15.26 | 0.03 | 57730.44140 | 14.62 | 0.04 |
| 57730.44206 | 17.20 | 0.04 | 57730.44274 | 15.79 | 0.04 | 57730.44327 | 15.27 | 0.03 | 57730.44384 | 14.70 | 0.04 |
| 57730.44445 | 17.11 | 0.04 | 57730.44511 | 15.81 | 0.04 | 57730.44561 | 15.24 | 0.03 | 57730.44616 | 14.63 | 0.04 |
| 57730.44680 | 17.13 | 0.04 | 57730.44745 | 15.80 | 0.04 | 57730.44797 | 15.30 | 0.03 | 57730.44851 | 14.66 | 0.04 |
| 57730.44915 | 17.16 | 0.04 | 57730.44981 | 15.74 | 0.04 | 57730.45033 | 15.34 | 0.03 | 57730.45090 | 14.65 | 0.04 |
| 57730.45154 | 17.13 | 0.04 | 57730.45218 | 15.84 | 0.04 | 57730.45273 | 15.26 | 0.03 | 57730.45327 | 14.67 | 0.04 |
| 57730.45391 | 17.20 | 0.04 | 57730.45456 | 15.80 | 0.04 | 57730.45510 | 15.30 | 0.03 | 57730.45565 | 14.63 | 0.04 |
| 57730.45629 | 17.22 | 0.04 | 57730.45693 | 15.76 | 0.04 | 57730.45746 | 15.29 | 0.03 | 57730.45801 | 14.64 | 0.04 |
| 57730.45866 | 17.10 | 0.04 | 57730.45930 | 15.82 | 0.04 | 57730.45984 | 15.27 | 0.03 | 57730.46039 | 14.60 | 0.04 |
| 57730.46102 | 17.19 | 0.04 | 57730.46167 | 15.71 | 0.04 | 57730.46220 | 15.31 | 0.03 | 57730.46276 | 14.57 | 0.04 |
| 57730.46339 | 17.17 | 0.04 | 57730.46403 | 15.81 | 0.04 | 57730.46457 | 15.26 | 0.03 | 57730.46512 | 14.62 | 0.04 |
| 57730.46576 | 17.20 | 0.04 | 57730.46638 | 15.79 | 0.04 | 57730.46692 | 15.27 | 0.03 | 57730.46747 | 14.62 | 0.04 |
| 57730.46810 | 16.99 | 0.04 | 57730.46874 | 15.80 | 0.04 | 57730.46927 | 15.29 | 0.03 | 57730.46983 | 14.66 | 0.04 |
| 57730.47047 | 17.02 | 0.04 | 57730.47111 | 15.76 | 0.04 | 57730.47165 | 15.30 | 0.03 | 57730.47221 | 14.63 | 0.04 |
| 57730.47286 | 17.17 | 0.04 | 57730.47350 | 15.82 | 0.04 | 57730.47404 | 15.29 | 0.03 | 57730.47458 | 14.65 | 0.04 |
| 57730.47521 | 17.06 | 0.04 | 57730.47585 | 15.85 | 0.04 | 57730.47639 | 15.24 | 0.03 | 57730.47694 | 14.66 | 0.04 |
| 57730.47757 | 17.21 | 0.04 | 57730.47821 | 15.80 | 0.04 | 57730.47876 | 15.24 | 0.03 | 57730.47931 | 14.67 | 0.04 |
| 57730.47995 | 17.17 | 0.04 | 57730.48059 | 15.80 | 0.04 | 57730.48113 | 15.30 | 0.03 | 57730.48169 | 14.65 | 0.04 |
| 57730.48234 | 17.10 | 0.04 | 57730.48298 | 15.83 | 0.04 | 57730.48353 | 15.32 | 0.03 | 57730.48408 | 14.61 | 0.04 |
| 57730.48472 | 17.18 | 0.04 | 57730.48539 | 15.80 | 0.04 | 57730.48592 | 15.30 | 0.03 | 57730.48648 | 14.71 | 0.04 |
| 57730.48713 | 17.13 | 0.04 | 57730.48779 | 15.78 | 0.04 | 57730.48833 | 15.26 | 0.03 | 57730.48889 | 14.64 | 0.04 |
| 57730.48953 | 17.22 | 0.04 | 57730.49019 | 15.78 | 0.04 | 57730.49073 | 15.25 | 0.03 | 57730.49128 | 14.65 | 0.04 |
| 57730.49192 | 17.18 | 0.04 | 57730.49257 | 15.82 | 0.04 | 57730.49312 | 15.30 | 0.03 | 57730.49367 | 14.65 | 0.04 |
| 57730.49431 | 17.22 | 0.04 | 57730.49497 | 15.77 | 0.04 | 57730.49552 | 15.31 | 0.03 | 57730.49610 | 14.66 | 0.04 |
| 57730.49674 | 17.10 | 0.04 | 57730.49742 | 15.78 | 0.04 | 57730.49795 | 15.29 | 0.03 | 57730.49853 | 14.64 | 0.04 |
| 57730.49988 | 17.17 | 0.04 | 57730.50055 | 15.78 | 0.04 | 57730.50107 | 15.28 | 0.03 | 57730.50165 | 14.65 | 0.04 |

| Gaia DR1 464387841322326272 ||||||||||||
|---|---|---|---|---|---|---|---|---|---|---|---|
| HJD (2400000 +) | B | Err. (±) | HJD (2400000 +) | V | Err. (±) | HJD (2400000 +) | $R_c$ | Err. (±) | HJD (2400000 +) | $I_c$ | Err. (±) |
| 57730.50230 | 17.16 | 0.04 | 57730.50297 | 15.81 | 0.04 | 57730.50350 | 15.29 | 0.03 | 57730.50408 | 14.61 | 0.04 |
| 57730.50473 | 17.16 | 0.04 | 57730.50538 | 15.84 | 0.04 | 57730.50592 | 15.29 | 0.03 | 57730.50650 | 14.66 | 0.04 |
| 57730.50715 | 17.15 | 0.04 | 57730.50784 | 15.79 | 0.04 | 57730.50838 | 15.30 | 0.03 | 57730.50891 | 14.65 | 0.04 |
| 57730.50957 | 17.24 | 0.04 | 57730.51022 | 15.76 | 0.04 | 57730.51076 | 15.24 | 0.03 | 57730.51135 | 14.63 | 0.04 |
| 57730.51199 | 17.13 | 0.04 | 57730.51265 | 15.76 | 0.04 | 57730.51319 | 15.29 | 0.03 | 57730.51375 | 14.62 | 0.04 |
| 57730.51442 | 17.20 | 0.04 | 57730.51507 | 15.82 | 0.04 | 57730.51563 | 15.21 | 0.03 | 57730.51618 | 14.63 | 0.04 |
| 57730.51684 | 17.10 | 0.04 | 57730.51750 | 15.78 | 0.04 | 57730.51804 | 15.29 | 0.03 | 57730.51861 | 14.66 | 0.04 |
| 57730.51928 | 17.10 | 0.04 | 57730.51992 | 15.83 | 0.04 | 57730.52047 | 15.29 | 0.03 | 57730.52104 | 14.64 | 0.04 |
| 57730.52172 | 17.02 | 0.04 | 57730.52239 | 15.80 | 0.04 | 57730.52294 | 15.28 | 0.03 | 57730.52351 | 14.69 | 0.04 |
| 57730.52419 | 17.12 | 0.04 | 57730.52484 | 15.81 | 0.04 | 57730.52539 | 15.30 | 0.03 | 57730.52599 | 14.67 | 0.04 |
| 57730.52665 | 17.29 | 0.04 | 57730.52732 | 15.81 | 0.04 |  |  |  |  |  |  |
| 57730.52877 | 17.19 | 0.04 | 57730.52937 | 15.79 | 0.04 | 57730.52986 | 15.31 | 0.03 | 57730.53037 | 14.63 | 0.04 |
| 57730.53097 | 17.26 | 0.04 | 57730.53158 | 15.78 | 0.04 | 57730.53207 | 15.29 | 0.03 | 57730.53258 | 14.69 | 0.04 |
| 57730.53317 | 17.09 | 0.04 | 57730.53379 | 15.78 | 0.04 | 57730.53428 | 15.29 | 0.03 | 57730.53479 | 14.69 | 0.04 |
| 57730.53540 | 17.14 | 0.04 | 57730.53601 | 15.83 | 0.04 | 57730.53650 | 15.30 | 0.03 | 57730.53702 | 14.66 | 0.04 |
| 57730.53762 | 17.10 | 0.04 | 57730.53823 | 15.82 | 0.04 | 57730.53872 | 15.26 | 0.03 | 57730.53925 | 14.63 | 0.04 |
| 57730.53985 | 17.10 | 0.04 | 57730.54045 | 15.88 | 0.04 | 57730.54096 | 15.24 | 0.03 | 57730.54147 | 14.66 | 0.04 |
| 57730.54208 | 17.12 | 0.04 | 57730.54268 | 15.81 | 0.04 | 57730.54318 | 15.35 | 0.03 | 57730.54371 | 14.64 | 0.04 |
| 57730.54431 | 17.02 | 0.04 | 57730.54492 | 15.88 | 0.04 | 57730.54542 | 15.27 | 0.03 | 57730.54594 | 14.69 | 0.04 |
| 57730.54656 | 17.27 | 0.04 | 57730.54717 | 15.76 | 0.04 | 57730.54767 | 15.24 | 0.03 | 57730.54820 | 14.61 | 0.04 |
| 57730.54882 | 17.19 | 0.04 | 57730.54943 | 15.79 | 0.04 | 57730.54994 | 15.25 | 0.03 | 57730.55046 | 14.67 | 0.04 |
| 57730.55107 | 17.26 | 0.04 | 57730.55171 | 15.78 | 0.04 | 57730.55221 | 15.29 | 0.03 | 57730.55274 | 14.66 | 0.04 |
| 57730.55336 | 17.14 | 0.04 | 57730.55398 | 15.82 | 0.04 | 57730.55449 | 15.31 | 0.03 | 57730.55502 | 14.63 | 0.04 |
| 57730.55564 | 17.25 | 0.04 | 57730.55626 | 15.82 | 0.04 | 57730.55678 | 15.33 | 0.03 | 57730.55730 | 14.73 | 0.04 |
| 57730.55794 | 17.16 | 0.04 | 57730.55855 | 15.83 | 0.04 | 57730.55907 | 15.34 | 0.03 | 57730.55961 | 14.64 | 0.04 |
| 57730.56022 | 17.15 | 0.04 | 57730.56085 | 15.81 | 0.04 | 57730.56138 | 15.26 | 0.03 | 57730.56190 | 14.65 | 0.04 |
| 57730.56252 | 17.25 | 0.04 | 57730.56316 | 15.83 | 0.04 | 57730.56366 | 15.28 | 0.03 | 57730.56420 | 14.74 | 0.04 |
| 57730.56483 | 17.25 | 0.04 | 57730.56546 | 15.74 | 0.04 | 57730.56599 | 15.24 | 0.03 | 57730.56653 | 14.65 | 0.04 |
| 57730.56716 | 17.19 | 0.04 | 57730.56778 | 15.89 | 0.04 | 57730.56832 | 15.37 | 0.03 | 57730.56884 | 14.65 | 0.04 |
| 57730.56947 | 17.18 | 0.04 | 57730.57012 | 15.83 | 0.04 | 57730.57063 | 15.27 | 0.03 | 57730.57117 | 14.68 | 0.04 |
| 57730.57180 | 17.21 | 0.04 | 57730.57244 | 15.80 | 0.04 | 57730.57297 | 15.29 | 0.03 | 57730.57352 | 14.65 | 0.04 |
| 57730.57415 | 17.20 | 0.04 | 57730.57479 | 15.84 | 0.04 | 57730.57531 | 15.32 | 0.03 | 57730.57586 | 14.60 | 0.04 |
| 57730.57649 | 17.15 | 0.04 | 57730.57712 | 15.81 | 0.04 | 57730.57765 | 15.26 | 0.03 | 57730.57818 | 14.63 | 0.04 |
| 57730.57881 | 17.15 | 0.04 | 57730.57946 | 15.83 | 0.04 | 57730.57998 | 15.33 | 0.03 | 57730.58052 | 14.64 | 0.04 |
| 57730.58115 | 17.26 | 0.04 | 57730.58180 | 15.84 | 0.04 | 57730.58232 | 15.29 | 0.03 | 57730.58287 | 14.66 | 0.04 |
| 57730.58349 | 17.17 | 0.04 | 57730.58414 | 15.78 | 0.04 | 57730.58467 | 15.29 | 0.03 | 57730.58522 | 14.65 | 0.04 |
| 57730.58585 | 17.19 | 0.04 | 57730.58648 | 15.82 | 0.04 | 57730.58702 | 15.34 | 0.03 | 57730.58756 | 14.69 | 0.04 |
| 57730.58818 | 17.20 | 0.04 | 57730.58882 | 15.82 | 0.04 | 57730.58935 | 15.35 | 0.03 | 57730.58990 | 14.67 | 0.04 |
| 57730.59052 | 17.09 | 0.04 | 57730.59118 | 15.85 | 0.04 | 57730.59171 | 15.29 | 0.03 | 57730.59225 | 14.67 | 0.04 |
| 57730.59289 | 17.10 | 0.04 | 57730.59352 | 15.86 | 0.04 | 57730.59405 | 15.24 | 0.03 | 57730.59458 | 14.66 | 0.04 |
| 57730.59521 | 17.15 | 0.04 | 57730.59586 | 15.88 | 0.04 | 57730.59637 | 15.27 | 0.03 | 57730.59692 | 14.74 | 0.04 |
| 57730.59778 | 17.26 | 0.04 | 57730.59843 | 15.79 | 0.04 | 57730.59894 | 15.29 | 0.03 | 57730.59949 | 14.68 | 0.04 |
| 57730.60013 | 17.09 | 0.04 | 57730.60078 | 15.81 | 0.04 | 57730.60129 | 15.29 | 0.03 | 57730.60184 | 14.64 | 0.04 |
| 57730.60248 | 17.22 | 0.04 | 57730.60310 | 15.79 | 0.04 | 57730.60365 | 15.31 | 0.03 | 57730.60419 | 14.63 | 0.04 |
| 57730.60483 | 17.18 | 0.04 | 57730.60554 | 15.77 | 0.04 | 57730.60608 | 15.27 | 0.03 | 57730.60666 | 14.64 | 0.04 |
| 57730.60731 | 17.24 | 0.04 | 57730.60800 | 15.76 | 0.04 | 57730.60855 | 15.35 | 0.03 | 57730.60910 | 14.68 | 0.04 |
| 57730.60973 | 17.19 | 0.04 | 57730.61037 | 15.78 | 0.04 | 57730.61091 | 15.27 | 0.03 | 57730.61146 | 14.65 | 0.04 |
| 57730.61211 | 17.28 | 0.04 | 57730.61277 | 15.83 | 0.04 | 57730.61330 | 15.28 | 0.03 | 57730.61385 | 14.62 | 0.04 |
| 57730.61462 | 17.30 | 0.04 | 57730.61525 | 15.74 | 0.04 | 57730.61579 | 15.30 | 0.03 | 57730.61634 | 14.67 | 0.04 |
| 57730.61697 | 17.24 | 0.04 | 57730.61763 | 15.77 | 0.04 | 57730.61817 | 15.27 | 0.03 | 57730.61873 | 14.64 | 0.04 |
| 57730.61937 | 17.29 | 0.04 | 57730.62000 | 15.82 | 0.04 | 57730.62054 | 15.28 | 0.03 | 57730.62110 | 14.66 | 0.04 |
| 57730.62174 | 17.06 | 0.04 | 57730.62238 | 15.82 | 0.04 | 57730.62292 | 15.34 | 0.03 | 57730.62346 | 14.69 | 0.04 |
| 57730.62410 | 17.23 | 0.04 | 57730.62475 | 15.82 | 0.04 | 57730.62528 | 15.30 | 0.03 | 57730.62583 | 14.67 | 0.04 |
| 57730.62646 | 17.15 | 0.04 | 57730.62712 | 15.82 | 0.04 | 57730.62764 | 15.29 | 0.03 | 57730.62819 | 14.67 | 0.04 |
| 57730.62883 | 17.21 | 0.04 | 57730.62947 | 15.82 | 0.04 | 57730.63001 | 15.33 | 0.03 | 57730.63056 | 14.64 | 0.04 |
| 57730.63119 | 17.27 | 0.04 | 57730.63182 | 15.79 | 0.04 | 57730.63236 | 15.28 | 0.03 | 57730.63289 | 14.67 | 0.04 |
| 57730.63352 | 17.25 | 0.04 | 57730.63417 | 15.78 | 0.04 | 57730.63470 | 15.42 | 0.03 | 57730.63525 | 14.66 | 0.04 |
| 57730.63589 | 17.06 | 0.04 | 57730.63654 | 15.92 | 0.04 | 57730.63707 | 15.28 | 0.03 | 57730.63763 | 14.61 | 0.04 |
| 57730.63855 | 17.28 | 0.04 | 57730.63914 | 15.81 | 0.04 | 57730.63964 | 15.28 | 0.03 | 57730.64015 | 14.67 | 0.04 |
| 57730.64075 | 17.19 | 0.04 | 57730.64134 | 15.83 | 0.04 | 57730.64183 | 15.22 | 0.03 | 57730.64235 | 14.59 | 0.04 |
| 57730.64296 | 17.33 | 0.04 | 57730.64356 | 15.78 | 0.04 | 57730.64406 | 15.30 | 0.03 | 57730.64457 | 14.62 | 0.04 |
| 57730.64540 | 17.30 | 0.04 | 57730.64601 | 15.83 | 0.04 | 57730.64651 | 15.24 | 0.03 | 57730.64703 | 14.62 | 0.04 |
| 57730.64801 | 17.22 | 0.04 | 57730.64862 | 15.77 | 0.04 | 57730.64911 | 15.34 | 0.03 | 57730.64964 | 14.64 | 0.04 |

| Gaia DR1 4643878413223262272 |||||||||
|---|---|---|---|---|---|---|---|---|
| HJD (2400000 +) | B-V | Err. (±) | HJD (2400000 +) | V-R | Err. (±) | HJD (2400000 +) | V-I | Err. (±) |
| 57730.34008 | 1.31 | 0.06 | 57730.34063 | 0.49 | 0.05 | 57730.34089 | 1.13 | 0.06 |
| 57730.34230 | 1.37 | 0.06 | 57730.34285 | 0.49 | 0.05 | 57730.34311 | 1.12 | 0.06 |
| 57730.34452 | 1.36 | 0.06 | 57730.34507 | 0.55 | 0.05 | 57730.34533 | 1.19 | 0.06 |
| 57730.34676 | 1.33 | 0.06 | 57730.34732 | 0.57 | 0.05 | 57730.34758 | 1.17 | 0.06 |
| 57730.34901 | 1.23 | 0.06 | 57730.34956 | 0.59 | 0.05 | 57730.34983 | 1.19 | 0.06 |
| 57730.35126 | 1.24 | 0.06 | 57730.35182 | 0.60 | 0.05 | 57730.35209 | 1.22 | 0.06 |
| 57730.35352 | 1.38 | 0.06 | 57730.35408 | 0.57 | 0.05 | 57730.35434 | 1.13 | 0.06 |
| 57730.35578 | 1.30 | 0.06 | 57730.35635 | 0.45 | 0.05 | 57730.35661 | 1.15 | 0.06 |
| 57730.35806 | 1.52 | 0.06 | 57730.35863 | 0.49 | 0.05 | 57730.35889 | 1.14 | 0.06 |
| 57730.36034 | 1.44 | 0.06 | 57730.36090 | 0.51 | 0.05 | 57730.36117 | 1.12 | 0.06 |
| 57730.36262 | 1.44 | 0.06 | 57730.36266 | 0.52 | 0.05 | 57730.36346 | 1.14 | 0.06 |
| 57730.36492 | 1.44 | 0.06 | 57730.36549 | 0.58 | 0.05 | 57730.36576 | 1.17 | 0.06 |
| 57730.36721 | 1.37 | 0.06 | 57730.36778 | 0.49 | 0.05 | 57730.36805 | 1.21 | 0.06 |
| 57730.36951 | 1.43 | 0.06 | 57730.37008 | 0.53 | 0.05 | 57730.37035 | 1.16 | 0.06 |
| 57730.37183 | 1.25 | 0.06 | 57730.37240 | 0.57 | 0.05 | 57730.37268 | 1.21 | 0.06 |
| 57730.37415 | 1.37 | 0.06 | 57730.37472 | 0.59 | 0.05 | 57730.37499 | 1.20 | 0.06 |
| 57730.37647 | 1.34 | 0.06 | 57730.37705 | 0.50 | 0.05 | 57730.37731 | 1.17 | 0.06 |
| 57730.37879 | 1.47 | 0.06 | 57730.37936 | 0.49 | 0.05 | 57730.37963 | 1.17 | 0.06 |
| 57730.38110 | 1.37 | 0.06 | 57730.38167 | 0.55 | 0.05 | 57730.38194 | 1.12 | 0.06 |
| 57730.38341 | 1.38 | 0.06 | 57730.38400 | 0.60 | 0.05 | 57730.38427 | 1.17 | 0.06 |
| 57730.38576 | 1.26 | 0.06 | 57730.38635 | 0.53 | 0.05 | 57730.38662 | 1.15 | 0.06 |
| 57730.38812 | 1.35 | 0.06 | 57730.38869 | 0.47 | 0.05 | 57730.38897 | 1.13 | 0.06 |
| 57730.39047 | 1.47 | 0.06 | 57730.39105 | 0.48 | 0.05 | 57730.39132 | 1.14 | 0.06 |
| 57730.39281 | 1.35 | 0.06 | 57730.39339 | 0.51 | 0.05 | 57730.39367 | 1.15 | 0.06 |
| 57730.39514 | 1.27 | 0.06 | 57730.39572 | 0.49 | 0.05 | 57730.39599 | 1.18 | 0.06 |
| 57730.39747 | 1.48 | 0.06 | 57730.39805 | 0.54 | 0.05 | 57730.39832 | 1.15 | 0.06 |
| 57730.39981 | 1.31 | 0.06 | 57730.40040 | 0.57 | 0.05 | 57730.40067 | 1.18 | 0.06 |
| 57730.40216 | 1.37 | 0.06 | 57730.40275 | 0.51 | 0.05 | 57730.40303 | 1.15 | 0.06 |
| 57730.40455 | 1.46 | 0.06 | 57730.40513 | 0.51 | 0.05 | 57730.40541 | 1.15 | 0.06 |
| 57730.40691 | 1.33 | 0.06 | 57730.40750 | 0.54 | 0.05 | 57730.40777 | 1.14 | 0.06 |
| 57730.42797 | 1.19 | 0.06 | 57730.42855 | 0.54 | 0.05 | 57730.42883 | 1.19 | 0.06 |
| 57730.43033 | 1.29 | 0.06 | 57730.43092 | 0.53 | 0.05 | 57730.43119 | 1.14 | 0.06 |
| 57730.43269 | 1.35 | 0.06 | 57730.43327 | 0.62 | 0.05 | 57730.43355 | 1.19 | 0.06 |
| 57730.43505 | 1.27 | 0.06 | 57730.43567 | 0.45 | 0.05 | 57730.43594 | 1.14 | 0.06 |
| 57730.43748 | 1.29 | 0.06 | 57730.43809 | 0.58 | 0.05 | 57730.43838 | 1.21 | 0.06 |
| 57730.43994 | 1.23 | 0.06 | 57730.44056 | 0.55 | 0.05 | 57730.44084 | 1.18 | 0.06 |
| 57730.44240 | 1.41 | 0.06 | 57730.44300 | 0.53 | 0.05 | 57730.44329 | 1.09 | 0.06 |
| 57730.44478 | 1.30 | 0.06 | 57730.44536 | 0.56 | 0.05 | 57730.44563 | 1.17 | 0.06 |
| 57730.44713 | 1.33 | 0.06 | 57730.44771 | 0.50 | 0.05 | 57730.44798 | 1.14 | 0.06 |
| 57730.44948 | 1.42 | 0.06 | 57730.45007 | 0.42 | 0.05 | 57730.45035 | 1.10 | 0.06 |
| 57730.45186 | 1.30 | 0.06 | 57730.45245 | 0.58 | 0.05 | 57730.45272 | 1.16 | 0.06 |
| 57730.45424 | 1.41 | 0.06 | 57730.45483 | 0.51 | 0.05 | 57730.45511 | 1.17 | 0.06 |
| 57730.45661 | 1.45 | 0.06 | 57730.45719 | 0.48 | 0.05 | 57730.45747 | 1.13 | 0.06 |
| 57730.45898 | 1.28 | 0.06 | 57730.45957 | 0.55 | 0.05 | 57730.45984 | 1.22 | 0.06 |
| 57730.46134 | 1.48 | 0.06 | 57730.46193 | 0.42 | 0.05 | 57730.46221 | 1.14 | 0.06 |
| 57730.46371 | 1.36 | 0.06 | 57730.46430 | 0.56 | 0.05 | 57730.46458 | 1.20 | 0.06 |
| 57730.46607 | 1.41 | 0.06 | 57730.46665 | 0.52 | 0.05 | 57730.46693 | 1.17 | 0.06 |
| 57730.46842 | 1.19 | 0.06 | 57730.46901 | 0.51 | 0.05 | 57730.46929 | 1.14 | 0.06 |
| 57730.47079 | 1.26 | 0.06 | 57730.47138 | 0.46 | 0.05 | 57730.47166 | 1.13 | 0.06 |
| 57730.47318 | 1.35 | 0.06 | 57730.47377 | 0.53 | 0.05 | 57730.47404 | 1.17 | 0.06 |
| 57730.47553 | 1.21 | 0.06 | 57730.47612 | 0.61 | 0.05 | 57730.47639 | 1.19 | 0.06 |
| 57730.47789 | 1.41 | 0.06 | 57730.47848 | 0.56 | 0.05 | 57730.47876 | 1.13 | 0.06 |
| 57730.48027 | 1.37 | 0.06 | 57730.48086 | 0.50 | 0.05 | 57730.48114 | 1.15 | 0.06 |
| 57730.48266 | 1.27 | 0.06 | 57730.48325 | 0.51 | 0.05 | 57730.48353 | 1.21 | 0.06 |
| 57730.48505 | 1.38 | 0.06 | 57730.48565 | 0.50 | 0.05 | 57730.48594 | 1.10 | 0.06 |
| 57730.48746 | 1.35 | 0.06 | 57730.48806 | 0.52 | 0.05 | 57730.48834 | 1.14 | 0.06 |
| 57730.48986 | 1.43 | 0.06 | 57730.49046 | 0.53 | 0.05 | 57730.49073 | 1.14 | 0.06 |
| 57730.49225 | 1.36 | 0.06 | 57730.49285 | 0.53 | 0.05 | 57730.49312 | 1.18 | 0.06 |
| 57730.49464 | 1.45 | 0.06 | 57730.49524 | 0.47 | 0.05 | 57730.49554 | 1.11 | 0.06 |
| 57730.49708 | 1.31 | 0.06 | 57730.49769 | 0.50 | 0.05 | 57730.49798 | 1.14 | 0.06 |
| 57730.50021 | 1.39 | 0.06 | 57730.50081 | 0.50 | 0.05 | 57730.50110 | 1.13 | 0.06 |
| 57730.50263 | 1.35 | 0.06 | 57730.50323 | 0.52 | 0.05 | 57730.50352 | 1.20 | 0.06 |
| 57730.50506 | 1.32 | 0.06 | 57730.50565 | 0.55 | 0.05 | 57730.50594 | 1.18 | 0.06 |
| 57730.50749 | 1.36 | 0.06 | 57730.50811 | 0.49 | 0.05 | 57730.50838 | 1.14 | 0.06 |

| Gaia DR1 464387841322326272 ||||||||
|---|---|---|---|---|---|---|---|---|
| HJD (2400000 +) | B-V | Err. (±) | HJD (2400000 +) | V-R | Err. (±) | HJD (2400000 +) | V-I | Err. (±) |
| 57730.50990 | 1.48 | 0.06 | 57730.51049 | 0.52 | 0.05 | 57730.51078 | 1.13 | 0.06 |
| 57730.51232 | 1.37 | 0.06 | 57730.51292 | 0.48 | 0.05 | 57730.51320 | 1.14 | 0.06 |
| 57730.51475 | 1.38 | 0.06 | 57730.51535 | 0.60 | 0.05 | 57730.51563 | 1.19 | 0.06 |
| 57730.51717 | 1.33 | 0.06 | 57730.51777 | 0.49 | 0.05 | 57730.51806 | 1.12 | 0.06 |
| 57730.51960 | 1.27 | 0.06 | 57730.52020 | 0.52 | 0.05 | 57730.52048 | 1.19 | 0.06 |
| 57730.52206 | 1.22 | 0.06 | 57730.52266 | 0.51 | 0.05 | 57730.52295 | 1.11 | 0.06 |
| 57730.52451 | 1.30 | 0.06 | 57730.52512 | 0.52 | 0.05 | 57730.52541 | 1.15 | 0.06 |
| 57730.52698 | 1.49 | 0.06 | | | | | | |
| 57730.52907 | 1.40 | 0.06 | 57730.52961 | 0.50 | 0.05 | 57730.52987 | 1.16 | 0.06 |
| 57730.53127 | 1.48 | 0.06 | 57730.53182 | 0.52 | 0.05 | 57730.53208 | 1.10 | 0.06 |
| 57730.53348 | 1.31 | 0.06 | 57730.53403 | 0.54 | 0.05 | 57730.53429 | 1.09 | 0.06 |
| 57730.53571 | 1.31 | 0.06 | 57730.53626 | 0.48 | 0.05 | 57730.53651 | 1.17 | 0.06 |
| 57730.53793 | 1.28 | 0.06 | 57730.53848 | 0.54 | 0.05 | 57730.53874 | 1.19 | 0.06 |
| 57730.54015 | 1.21 | 0.06 | 57730.54071 | 0.63 | 0.05 | 57730.54096 | 1.22 | 0.06 |
| 57730.54238 | 1.31 | 0.06 | 57730.54293 | 0.56 | 0.05 | 57730.54319 | 1.17 | 0.06 |
| 57730.54462 | 1.14 | 0.06 | 57730.54517 | 0.58 | 0.05 | 57730.54543 | 1.18 | 0.06 |
| 57730.54686 | 1.51 | 0.06 | 57730.54742 | 0.47 | 0.05 | 57730.54768 | 1.15 | 0.06 |
| 57730.54913 | 1.40 | 0.06 | 57730.54969 | 0.47 | 0.05 | 57730.54995 | 1.12 | 0.06 |
| 57730.55139 | 1.48 | 0.06 | 57730.55196 | 0.45 | 0.05 | 57730.55223 | 1.12 | 0.06 |
| 57730.55367 | 1.33 | 0.06 | 57730.55423 | 0.56 | 0.05 | 57730.55450 | 1.18 | 0.06 |
| 57730.55595 | 1.44 | 0.06 | 57730.55652 | 0.54 | 0.05 | 57730.55678 | 1.09 | 0.06 |
| 57730.55825 | 1.33 | 0.06 | 57730.55881 | 0.58 | 0.05 | 57730.55908 | 1.19 | 0.06 |
| 57730.56053 | 1.35 | 0.06 | 57730.56112 | 0.44 | 0.05 | 57730.56137 | 1.15 | 0.06 |
| 57730.56284 | 1.42 | 0.06 | 57730.56341 | 0.56 | 0.05 | 57730.56368 | 1.10 | 0.06 |
| 57730.56515 | 1.51 | 0.06 | 57730.56573 | 0.47 | 0.05 | 57730.56600 | 1.10 | 0.06 |
| 57730.56747 | 1.30 | 0.06 | 57730.56805 | 0.57 | 0.05 | 57730.56831 | 1.23 | 0.06 |
| 57730.56980 | 1.35 | 0.06 | 57730.57037 | 0.57 | 0.05 | 57730.57064 | 1.15 | 0.06 |
| 57730.57212 | 1.41 | 0.06 | 57730.57271 | 0.48 | 0.05 | 57730.57298 | 1.15 | 0.06 |
| 57730.57447 | 1.36 | 0.06 | 57730.57505 | 0.55 | 0.05 | 57730.57533 | 1.24 | 0.06 |
| 57730.57680 | 1.33 | 0.06 | 57730.57738 | 0.52 | 0.05 | 57730.57765 | 1.18 | 0.06 |
| 57730.57914 | 1.32 | 0.06 | 57730.57972 | 0.49 | 0.05 | 57730.57999 | 1.19 | 0.06 |
| 57730.58147 | 1.42 | 0.06 | 57730.58206 | 0.49 | 0.05 | 57730.58233 | 1.18 | 0.06 |
| 57730.58382 | 1.38 | 0.06 | 57730.58441 | 0.50 | 0.05 | 57730.58468 | 1.14 | 0.06 |
| 57730.58617 | 1.37 | 0.06 | 57730.58675 | 0.58 | 0.05 | 57730.58702 | 1.13 | 0.06 |
| 57730.58850 | 1.37 | 0.06 | 57730.58908 | 0.56 | 0.05 | 57730.58936 | 1.15 | 0.06 |
| 57730.59085 | 1.24 | 0.06 | 57730.59144 | 0.56 | 0.05 | 57730.59171 | 1.18 | 0.06 |
| 57730.59320 | 1.24 | 0.06 | 57730.59378 | 0.56 | 0.05 | 57730.59405 | 1.19 | 0.06 |
| 57730.59554 | 1.27 | 0.06 | 57730.59611 | 0.56 | 0.05 | 57730.59639 | 1.14 | 0.06 |
| 57730.59811 | 1.47 | 0.06 | 57730.59868 | 0.52 | 0.05 | 57730.59896 | 1.12 | 0.06 |
| 57730.60045 | 1.28 | 0.06 | 57730.60103 | 0.47 | 0.05 | 57730.60131 | 1.17 | 0.06 |
| 57730.60279 | 1.43 | 0.06 | 57730.60338 | 0.53 | 0.05 | 57730.60365 | 1.16 | 0.06 |
| 57730.60519 | 1.41 | 0.06 | 57730.60581 | 0.49 | 0.05 | 57730.60610 | 1.13 | 0.06 |
| 57730.60765 | 1.48 | 0.06 | 57730.60827 | 0.47 | 0.05 | 57730.60855 | 1.09 | 0.06 |
| 57730.61005 | 1.41 | 0.06 | 57730.61064 | 0.52 | 0.05 | 57730.61091 | 1.14 | 0.06 |
| 57730.61244 | 1.45 | 0.06 | 57730.61304 | 0.56 | 0.05 | 57730.61331 | 1.21 | 0.06 |
| 57730.61494 | 1.57 | 0.06 | 57730.61552 | 0.41 | 0.05 | 57730.61580 | 1.08 | 0.06 |
| 57730.61730 | 1.47 | 0.06 | 57730.61790 | 0.48 | 0.05 | 57730.61818 | 1.14 | 0.06 |
| 57730.61968 | 1.47 | 0.06 | 57730.62027 | 0.53 | 0.05 | 57730.62055 | 1.16 | 0.06 |
| 57730.62206 | 1.24 | 0.06 | 57730.62265 | 0.49 | 0.05 | 57730.62292 | 1.12 | 0.06 |
| 57730.62442 | 1.41 | 0.06 | 57730.62501 | 0.54 | 0.05 | 57730.62529 | 1.16 | 0.06 |
| 57730.62679 | 1.33 | 0.06 | 57730.62738 | 0.40 | 0.05 | 57730.62766 | 1.14 | 0.06 |
| 57730.62915 | 1.39 | 0.06 | 57730.62974 | 0.54 | 0.05 | 57730.63001 | 1.18 | 0.06 |
| 57730.63150 | 1.48 | 0.06 | 57730.63209 | 0.52 | 0.05 | 57730.63236 | 1.13 | 0.06 |
| 57730.63385 | 1.47 | 0.06 | 57730.63444 | 0.56 | 0.05 | 57730.63471 | 1.13 | 0.06 |
| 57730.63621 | 1.14 | 0.06 | 57730.63680 | 0.61 | 0.05 | 57730.63708 | 1.30 | 0.06 |
| 57730.63885 | 1.47 | 0.06 | 57730.63939 | 0.57 | 0.05 | 57730.63965 | 1.15 | 0.06 |
| 57730.64104 | 1.36 | 0.06 | 57730.64159 | 0.50 | 0.05 | 57730.64185 | 1.24 | 0.06 |
| 57730.64326 | 1.56 | 0.06 | 57730.64381 | 0.56 | 0.05 | 57730.64407 | 1.16 | 0.06 |
| 57730.64570 | 1.48 | 0.06 | 57730.64626 | 0.50 | 0.05 | 57730.64652 | 1.21 | 0.06 |
| 57730.64832 | 1.46 | 0.06 | 57730.64887 | 0.45 | 0.06 | 57730.64913 | 1.14 | 0.06 |